\documentclass[12pt]{article}
\usepackage[top=2.5cm,bottom=3.5cm,left=2.5cm,right=2.5cm,footskip=1.5cm]{geometry}
\usepackage{amsmath,amssymb}
\usepackage{amsthm}
\usepackage{amscd}
\usepackage[all]{xy}
\usepackage{mathrsfs}
\usepackage{xcolor}
\numberwithin{equation}{section}
\def \nn{\nonumber\\}
\newcommand{\co}{{\rm c}}
\newcommand{\Po}{{\text{P}}}%
\newcommand{\na}{\nabla}
\renewcommand{\b}{\bar}
\renewcommand{\d}{\dot}
\newcommand{\pd}[2]{\frac{\partial{#1}}{\partial{#2}}}
\newcommand{\sr}{\sqrt}
\newcommand{\df}{\dfrac}
\newcommand{\ul}{\underline}
\newcommand{\der}{\partial}
\renewcommand{\(}{\left(}
\renewcommand{\)}{\right)}
\newcommand{\scr}{\mathscr}
\renewcommand{\cal}{\mathcal}
\newcommand{\dg}{\dagger}
\newcommand{\vs}[1]{\vspace{#1 mm}}

\newdimen\Tdim
\newdimen\Ddim
\def\Tspan#1{{\setbox0=\hbox{$#1$}%
\Tdim\ht0\advance\Tdim\dp0\advance\Tdim.7ex\Ddim\dp0\advance\Ddim.4ex\rule[-\Ddim]{0pt}{\Tdim}\box0}}

\newcommand{\Sqrbra}[4]{\raise-#4\hbox{
\rule{#3}{#1}\hskip-#3
\rule{#2}{#3}\hskip-#2
\rule[#1]{#2}{#3}}
\hskip-#2
}
\newcommand{\Sqrket}[5]{\raise-#4\hbox{
\rule{#2}{#3}\hskip-#2
\rule[#1]{#2}{#3}
\rule[#1]{#3}{#3}
\hskip-#5
\rule{#3}{#1}}}

\def\cfbra{\Sqrbra{14pt}{4.1pt}{0.45pt}{3.9pt}
\kern-.61pt\big(}
\def\Bigbra{\Sqrbra{21pt}{5.5pt}{0.5pt}{7.8pt}
\kern-.7pt\Big(}
\def\biggbra{\Sqrbra{28.3pt}{7pt}{0.55pt}{11.5pt}
\kern-0.85pt\bigg(}
\def\cfket{
\big) \kern-5.45pt\Sqrket{14pt}{4.05pt}{0.45pt}{3.9pt}{4.54pt}}
\def\Bigket{
\Big) \kern-6.5pt\Sqrket{21pt}{5.45pt}{0.5pt}{7.8pt}{4.8pt}}
\def\biggket{
\bigg) \kern-7.8pt\Sqrket{28.3pt}{6pt}{0.55pt}{11.5pt}{4.5pt}}

\begin{document}
\baselineskip=14.5pt
\begin{titlepage}
\hfill MISC-2016-05
\vspace*{20pt}
\begin{center}
{\Large\bf Superspace Gauge Fixing in
Yang-Mills Matter\\[3mm] Coupled Conformal Supergravity%
}\vs{14}

{\large
Taichiro Kugo$^a$, 
Ryo Yokokura$^b$ 
and Koichi Yoshioka$^c$} \\
\vs{6}

$^a${\it Department of Physics and
Maskawa Institute for Science and Culture,}\\
{\it Kyoto Sangyo University, Kyoto 603-8555, Japan}\\
$^b${\it Department of Physics,
Keio  University, Yokohama 223-8522, Japan}\\
$^c${\it Osaka University of Pharmaceutical Sciences, Takatsuki 
569-1094, Japan}

\vs{15}
{\bf Abstract}

\end{center}

\noindent
In $D=4$, $\cal{N}=1$ conformal superspace, 
the Yang-Mills matter coupled supergravity system is constructed where 
the Yang-Mills gauge interaction is introduced by 
extending the superconformal group to include the 
K\"ahler isometry group of chiral matter fields. 
There are two gauge-fixing procedures 
to get to the component Poincar\'e supergravity: 
one via the superconformal component formalism and the other via the
Poincar\'e superspace formalism. 
These two types of superconformal gauge-fixing conditions are analyzed
in detail and their correspondence is clarified. 
\end{titlepage}

\section{Introduction}

In our previous paper~\cite{bib: KYY} we have demonstrated the
equivalence between the conventional component
approach~\cite{bib:KTVN}-\cite{bib: KKLVP} as known as the
superconformal tensor calculus and the recent superspace
approach~\cite{bib: B1,bib: B2} to $D=4$, $\cal{N}=1$ conformal
supergravity (SUGRA). The detailed correspondences between two
approaches were explicitly shown for superconformal gauge fields,
curvatures and curvature constraints, general conformal multiplets and
their transformation laws. We also briefly discussed the
superconformal gauge fixing which leads to the Poincar\'e SUGRA.

The previous analysis was not sufficient in two points. One is that we
have confined ourselves to the matter coupled SUGRA system in which
there is no Yang-Mills (YM) gauge field of internal symmetry.
Such YM interactions 
should generally be introduced by gauging the isometry of the K\"ahler 
manifold of chiral matter fields, which requires some extra work 
in conformal superspace approach. 
This way of including YM interactions in superspace has essentially 
been known for the super-Poincar\'e 
case~\cite{Bagger:1982ab}-\cite{bib: BG2}, but not for the 
superconformal case.%
\footnote{For some special cases, 
YM interactions in conformal superspace were discussed, e.g.,
for the linear compensator~\cite{bib: B2} and for the
pure Fayet-Iliopoulos $U(1)$ system~\cite{bib: BFI}.}  
As we will show in section 2, that is a simpler task
in conformal superspace than in the super-Poincar\'e case, thanks 
to the simplicity of the algebra for covariant derivatives. 

The isometry transformation of the K\"ahler manifold does not necessarily 
leave the K\"ahler potential invariant but induces the so-called 
K\"ahler transformation, i.e., a shift by holomorphic and 
anti-holomorphic functions. 
In this case, the chiral compensator is also transformed under 
the YM gauge transformation~\cite{{bib: FGKVP},{bib: KKLVP}}. 
In section 3, we discuss the superconformal gauge fixing in superspace
for the YM matter coupled SUGRA system. 
Two types of gauge fixing are studied for realizing 
the canonically normalized Einstein-Hilbert (EH) and Rarita-Schwinger (RS) 
terms: one is applicable for non-vanishing superpotential 
and the other is independent of the superpotential term.
For the former case, the superconformal gauge-fixing condition is YM gauge 
invariant so that the previous results in Ref.~\cite{bib: KYY} almost
hold. For the latter case, however, the gauge-fixing
condition is not YM gauge invariant, which causes a modification of
the YM gauge transformation 
and the covariant derivatives in the resultant Poincar\'e superspace. 

Another insufficient point is that, while we have seen the correspondence 
between the two approaches on many quantities, we have not directly touched 
on the fact that the superspace formalism has much more gauge freedom
and gauge fields than the component formalism. In superspace, 
the gauge fields and the gauge transformation parameters are superfields
with higher $\theta$ components, and further
the gauge superfields contain the spinor-indexed components, all of
which do not appear in component approach. 
In section~4, we discuss how these extra 
degrees of freedom are fixed in order to have the component formalism. 
We explicitly give all the necessary gauge-fixing conditions 
component-field-wise
so as to fix the higher $\theta$ components of superfield gauge
transformation parameters. This gauge fixing from conformal superspace to 
superconformal component approach is depicted as the route I
in Fig.~\ref{fig:aim}. We show that the 
resultant theory after the gauge fixing 
agrees with the superconformal tensor calculus, that is, 
all the extra gauge fields are 
fixed to zero or reduced to the known quantities in component approach.

On the other hand, the gauge fixing discussed in section 3 
corresponds to the route II in Fig.~\ref{fig:aim}, where the gauge fixing 
is given superfield-wise to go down to the Poincar\'e superspace. 
The obtained Poincar\'e superspace formulation still has the
gauge invariance with superfield gauge transformation parameters and 
their higher $\theta$ gauge invariance
should be fixed to have the component Poincar\'e SUGRA\@. 
Such gauge fixing, the route III in Fig.~\ref{fig:aim}, can be done 
in the same way as the superconformal case (the route I) that
is clear from the discussion in section~4.

In section 5, we clarify in more detail how some type of superspace
gauge fixing (the route II) corresponds to the so-called improved
superconformal gauge in component approach (the route IV). It is
noticed that there is a small puzzle: 
The component formulation seems to be obtained from the superspace one
by setting the spinor-indexed components of gauge fields to zero. 
Nevertheless, in the Poincar\'e SUGRA obtained via the route 
II$+$III, non-vanishing $A$- and $K_A$-gauge fields 
with spinor indices remain and show up in the $A$- and $K_A$-gauge 
transformation parts in the Poincar\'e supersymmetry transformation. 
We give an answer to this in view of the resetting of gauge-fixing conditions.

\def\ffbox#1{\fbox{\rule[-1ex]{0pt}{3ex}\rule{1eM}{0pt}#1\rule{1eM}{0pt}}}
\begin{figure}[t]
\[
\xymatrix
{
\hbox{\underline{superspace approach}} &&&&&& 
\hbox{\underline{component approach}} \\
\ffbox{\txt{Conformal \\ superspace}}
\ar[rrrrrr]_{\hbox{I}}^{\hbox{\small fixing higher $\theta$ comps.~of $\,\xi^{\cal{A}}$}}
\ar[dd]_{\hbox{II}\ }^{\txt{\small fixing superfield gauge freedom \\$\xi(D)$, $\xi(A)$, $\xi(K)^A$}}
&&&&&&
\ffbox{\txt{superconformal \\tensor calculus}}
\ar[dd]_{\hbox{IV}\ }^{\txt{\small fixing gauge freedom \\ for $D,\,A,\,K_A$ }}
\\ 
&&&&&&
\\
\ffbox{\txt{Poincar\'e \\ superspace}}
\ar[rrrrrr]
^{\hbox{III}}
_{\txt{\small fixing higher $\theta$ comps.~of\\ 
\small \ $\xi(P)^A$, $\xi(M)^{ab}$, $\xi^{(a)}$} }
&&&&&&
\ffbox{\hbox{Poincar\'e SUGRA}}
}  
\]
\caption{Relations among four SUGRA formulations: superspace and
component approaches possessing superconformal and
super-Poincar\'e gauge symmetries.  
There are two routes I$+$IV and II$+$III for getting to the Poincar\'e
SUGRA from the conformal superspace formulation.
The symbol $\xi$ denotes the gauge transformation parameters of the
superconformal and internal symmetries (see the text for details). 
}\bigskip
\label{fig:aim}
\end{figure}
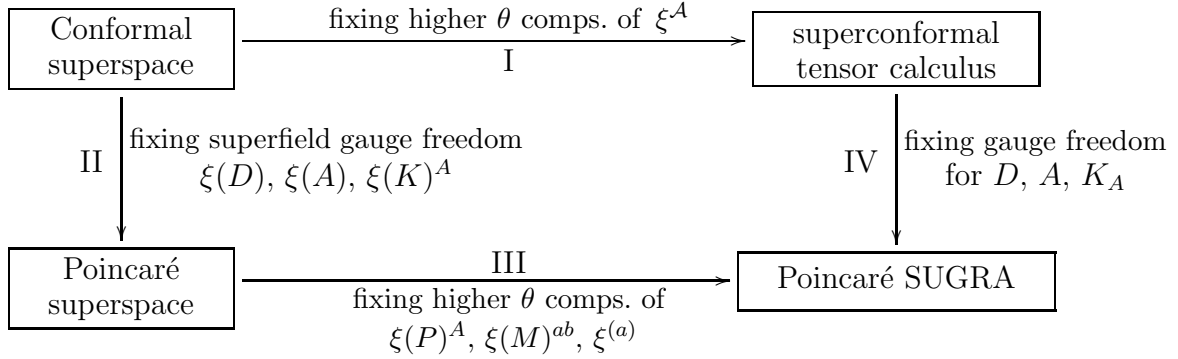

\section{YM matter coupled SUGRA in conformal superspace}
\label{sec: scym}

In this section, we introduce the YM system in conformal superspace.
The YM system is coupled to matter fields as an internal gauge symmetry.
The internal gauge symmetry is the 
isometry of the K\"ahler manifold spanned by chiral matter fields, which 
isometry is generally given by nonlinear transformation.
We use the so-called covariant approach in which one 
extends the superconformal covariant derivatives to be also
covariant under the internal gauge symmetry~\cite{{bib: SS},{bib: BG2}}. 
One advantage of the covariant approach is that 
the gauge transformation parameters are taken to be real (general)
superfields, and hence the internal gauge symmetry is made manifest.

Then we present two types of superconformal gauge-fixing conditions which 
realize the canonically normalized EH and RS terms.
One gives a real gravitino mass parameter and is
adopted only when the superpotential does not vanish.
The other can be imposed even when the superpotential vanishes but
leads to a complex gravitino mass.

We consider the Lie algebra of a compact Lie group $\scr{G}$ as an
internal symmetry.
The elements of the Lie algebra are denoted by $X_{(a)}$
where $(a)=1,2,...,\dim \scr{G}$. Their commutation relation is
\begin{equation}
  [X_{(a)},X_{(b)}]=-{f_{(a)(b)}}^{(c)}X_{(c)}, 
\label{eq:Xacom}
\end{equation}
where 
${f_{(a)(b)}}^{(c)}$ is the structure constant of the Lie algebra.
Since the spacetime symmetry and the internal symmetry are mutually
independent, the elements of these two algebras commute with each other.
We introduce the real gauge superfield $\cal{A}^{(a)}_M$ for the 
internal symmetry. 
In the following, we deal with the superconformal and internal
symmetries simultaneously and denote all the elements collectively by
$X_{\cal{A}}$. The gauge superfields and parameter superfields are 
denoted as
\begin{equation}
\begin{split}
 h_M{}^{\cal{A}}X_{\cal{A}}
&\,=\, {E_{M}}^A P_A+\df{1}{2}{\phi_M}^{ba}M_{ab}+B_M D+A_M A+{f_M}^AK_A
+\cal{A}_M^{(a)}X_{(a)},\ \ \ \  \\
 {\xi}^{\cal{A}}X_{\cal{A}}
&\,=\, {\xi}(P)^AP_A+\df{1}{2}{\xi}(M)^{ba}M_{ab}+\xi(D) D+\xi(A)
A+{\xi}(K)^{A}K_A +\xi^{(a)}X_{(a)}.
\end{split}
\end{equation}
Here, $E_M{}^A$ is the vielbein superfield corresponding to the
translation and supersymmetry generators
$P_A=(P_a, Q_\alpha,\b{Q}^{\d\alpha})$, 
$\phi_M{}^{ba}$ is the spin connection corresponding to the Lorentz
generator $M_{ab}$, and $B_M$, $A_M$, $f_M{}^A$ are the gauge
superfields corresponding to the dilatation $D$, the $U(1)$ chiral
transformation $A$, the conformal boost and its supersymmetry 
$K_A=(K_a,S_\alpha,\b{S}^{\d\alpha})$, respectively.
The gauge transformation parameter $\xi^{\cal{A}}$ is a real superfield.
The deformed $P_A$ transformation and covariant derivatives are
defined by the general coordinate transformation $\delta_{\text{GC}}$ 
and the gauge transformation $\delta_G$ as
\begin{equation}
\begin{split}
\delta_G(\xi(P)^AP_A)
&=\delta_{\text{GC}}(\xi^AE_A{}^M)
-\delta_G(\xi^BE_B{}^Mh_M{}^{\cal{A}'}X_{\cal{A}'}),\\
\na_M
&=\der_M -\df{1}{2}{\phi_M}^{ba}M_{ab}-B_M D-A_M A-{f_M}^AK_A
-\cal{A}_M^{(a)}X_{(a)}.
\end{split}
\end{equation}
Here we set the parameter $\xi(P)^A$ to be field-independent. The
symbol $X_{\cal{A}'}$ means the generators other than $P_A$.
The gauge superfield $\cal{A}^{(a)}_M$ is
transformed under the superconformal and internal symmetries as 
\begin{equation}
 \delta(\xi^{\cal{A}}X_{\cal{A}}) \cal{A}^{(a)}_M
= \der_M \xi^{(a)}+\cal{A}^{(b)}_M \xi^{(c)}
 {f_{(c)(b)}}^{(a)}+{E_M}^B\xi(P)^{C}\cal{F}_{CB}^{(a)},
\end{equation}
where $\cal{F}_{MN}^{(a)}$ is the curvature superfield 
for the internal symmetry:
\begin{equation}
\cal{F}^{(a)}_{MN}=\der_M \cal{A}^{(a)}_N-\der_N \cal{A}^{(a)}_M 
-\cal{A}^{(b)}_N\cal{A}_M^{(c)}{f_{(c)(b)}}^{(a)}.
\end{equation}
Similarly to the case without YM, we impose the curvature constraints 
$\{\na_\alpha, \na_\beta\}=0$, $\{\b\na_{\d\alpha},\b\na_{\d\beta}\}=0$, and
$\{\na_\alpha,\b\na_{\d\beta}\}=-2i\na_{\alpha\d\beta}$, which 
implies $\cal{F}^{(a)}_{\alpha\beta}=\cal{F}^{(a)}_{\d\alpha\d\beta}=
\cal{F}^{(a)}_{\alpha\d\beta}=0$ in the YM part. 
Solving the Bianchi identities under these constraints, we find 
that the curvatures $R_{\ul{\alpha} b}$
and $R_{ab}$ can be expressed by a single ``gaugino'' superfield 
$\cal{W}_{\ul{\alpha}}$ as follows ($\ul{\alpha}=(\alpha,\dot{\alpha})$)
\begin{gather}
R_{\alpha,\beta\d\gamma} = -[\na_\alpha,\na_{\beta\d\gamma}]
=2i\epsilon_{\alpha\beta}\cal{W}_{\d\gamma},  \qquad
R_{\d\alpha,\d\beta\gamma} = -[\b\na_{\d\alpha},\na_{\d\beta\gamma}]
=2i\epsilon_{\d\alpha\d\beta}\cal{W}_\gamma, 
\label{eq:Rspinorvector}  \\[1mm]
R_{\alpha\d\alpha,\beta\d\beta} 
= -[\na_{\alpha\d\alpha},\na_{\beta\d\beta}]
= -\epsilon_{\d\alpha\d\beta}
\{\na_{(\alpha},\cal{W}_{\beta)}\} -\epsilon_{\alpha\beta}
\{\b\na_{(\d\alpha},\cal{W}_{\d\beta)}\}, 
\label{eq:RvvW}
\end{gather}
and $\cal{W}_{\ul{\alpha}}$ contains the
YM gaugino superfield $\cal{W}^{(a)}_{\ul{\alpha}}$:
\begin{align}
 \cal{W}_\alpha&=
(\epsilon\sigma^{bc})^{\beta\gamma}W_{\alpha\beta\gamma}M_{cb}
+\df{1}{2}\nabla^{\gamma}{W_{\gamma\alpha}}^{\beta}S_\beta 
-\df{1}{2}\nabla^{\gamma\dot{\beta}}{W_{\gamma\alpha}}^{\beta}
K_{\beta\dot{\beta}}
+\cal{W}^{(a)}_\alpha X_{(a)}, \\
 {\cal{W}}^{\d\alpha} &=
(\b\sigma^{bc}\epsilon)^{\d\gamma\d\beta}{W}^{\d\alpha}_{
\hphantom{\d\alpha}\d\beta\d\gamma}M_{cb}
-\df{1}{2}\bar\nabla_{\d\gamma}{W}^{\d\gamma\d\alpha}_{
\hphantom{\d\gamma\d\alpha}\d\beta}\b{S}^{\d\beta}
-\df{1}{2}\nabla^{\d\gamma\beta}{W}_{\d\gamma}^{
\hphantom{{\d\gamma}}\d\alpha\d\beta}K_{\beta\d\beta}
+{\cal{W}}^{(a)\d\alpha} X_{(a)}.
\end{align}
Eq.~(\ref{eq:Rspinorvector}) implies in the YM part
\begin{align}
&\cal{F}_{\alpha,\beta\d\gamma}^{(a)}=2i \epsilon_{\alpha\beta}
  {\cal{W}}^{(a)}_{\d\gamma},\qquad
 \cal{F}^{(a)}_{\d\alpha,\gamma\d\beta}=
2i\epsilon_{\d\alpha\d\beta}\cal{W}^{(a)}_{\gamma}, \\
& \quad \cal{F}^{(a)}_{\alpha\dot{\alpha},\beta\dot{\beta}}
=-\epsilon_{\dot{\alpha}\dot{\beta}}
\nabla_{(\alpha}\cal{W}^{(a)}_{\beta)}
-\epsilon_{\alpha\beta}
{\bar\nabla}_{(\dot{\alpha}}{\cal{W}}^{(a)}_{\dot{\beta})}.
\end{align}
The gaugino superfield $\cal{W}_{\ul{\alpha}}$, particularly
$\cal{W}^{(a)}_{\ul{\alpha}}$ is found to satisfy the following
superconformal property from the Bianchi/Jacobi identities, 
\begin{gather}
\na^\alpha\cal{W}_\alpha^{(a)}=
\bar\na_{\d\alpha}{\cal{W}}^{\d\alpha(a)},\\ 
\bar\na_{\d\alpha} \cal{W}^{(a)}_\alpha=0, \ \ 
D\cal{W}_\alpha^{(a)}
=\df{3}{2}\cal{W}_\alpha^{(a)}, \ \  
A\cal{W}_\alpha^{(a)}
=i\cal{W}_{\alpha}^{(a)}, \ \
K_A\cal{W}_\alpha^{(a)}=0.
\label{eq: Wa}
\end{gather}
That is, $\cal{W}^{(a)}_\alpha$ is a covariantly chiral and
primary superfield carrying the Weyl weight $\Delta$ and the chiral
weight $w$ with $(\Delta,w)=(3/2,1)$. Note 
that $(\cal{W}^{(a)}_\alpha)^{\dg}=-\cal{W}^{(a)}_{\d\alpha}$,
and our $\cal{A}^{(a)}_M$ and $\cal{W}_{\alpha}^{(a)}$
are equivalent to $-i\cal{A}^{(r)}_M$ and $-i\cal{W}^{(r)}_\alpha$
in \cite{bib: BG2}, respectively.

The coupling of YM to matter superfields in conformal superspace 
can be discussed in a similar way as in the component 
approach \cite{bib: KKLVP}.
The matter primary superfields $\Phi^i$ ($i=1,2,\ldots,n$) have
the Weyl and chiral weights $(\Delta,w)=(0,0)$, and 
are covariantly chiral
with respect to the superconformal and internal symmetries:
\begin{equation}
  \bar\na^{\d\alpha}\Phi^i=0.
\end{equation}
The internal symmetry preserves the metric of the K\"ahler manifold 
spanned by chiral matter fields and their
conjugates. The metric of the manifold is written by
\begin{equation}
 g_{ij^*}=\pd{^2 K}{\Phi^i\der\b\Phi^{j^*}},
\end{equation}
where the K\"ahler potential $K$ is 
a real function of $\Phi^i$ and $\bar\Phi^{i^*}$. 

The generator $X_{(a)}$ acts on the matter superfields as a vector 
superfield $V_{(a)}$, which
can be decomposed into the holomorphic part $V^-_{(a)}$ 
and anti-holomorphic one $V^+_{(a)}$ :
\begin{equation}
V_{(a)}= V^-_{(a)}+V^+_{(a)},
\qquad 
V^-_{(a)}=V^i_{(a)}(\Phi)\pd{}{\Phi^i},
\quad 
V^+_{(a)}=\b{V}^{i^*}_{(a)}(\b\Phi)\pd{}{\b\Phi^{i^*}}.
\end{equation}
Thus, $X_{(a)}$ acts as 
\begin{equation}
 X_{(a)}\Phi^i=V_{(a)}^i(\Phi),
\qquad
X_{(a)}\b\Phi^{i^*}=\b{V}_{(a)}^{i^*}(\b\Phi),
\end{equation}
and preserves the metric of the manifold, which means 
$V_{(a)}$ is the Killing vector:
\begin{equation}
V_{(a)}^i\pd{g_{jk^*}}{\Phi^i}+\pd{V_{(a)}^i}{\Phi^j}g_{ik^*}+
\b{V}_{(a)}^{i^*}\pd{g_{jk^*}}{\b\Phi^{i^*}}
+\pd{\b{V}_{(a)}^{i^*}}{\b\Phi^{k^*}}g_{ji^*}=0.
\end{equation}
Solving this equation, the action of $V_{(a)}^{\pm}$ on the K\"ahler
potential is found to be expressed as
\begin{equation}
 V_{(a)}^iK_i=F_{(a)}-iJ_{(a)},\qquad
  \b{V}_{(a)}^{i^*}K_{i^*}=\b{F}_{(a)}+iJ_{(a)},
\label{eq: FJ}
\end{equation} 
where $K_i=\partial K/\partial\Phi^i$ and $K_{i^*}=
\partial K/\partial\b{\Phi}^{i^*}$. On the RHS, $F_{(a)}$ is 
a holomorphic function and $J_{(a)}$ is a real function called the
Killing potential or moment map.

The superspace action of YM matter coupled conformal SUGRA is given by
\begin{equation}
\begin{split}
 S&=-3\int d^4x d^4\theta\, E\, \Phi^\co\b\Phi^\co e^{-K/3}  \\
&\qquad + \(
\int d^4xd^2\theta\,\cal{E}\, (\Phi^\co)^3W(\Phi)
-\df{1}{4}\int d^4xd^2\theta\,\cal{E}\, H_{(a)(b)}(\Phi)
\cal{W}^{\alpha(a)}\cal{W}_\alpha^{(b)}+\text{h.c.} \).
\label{eq:action}
\end{split}
\end{equation}
Here the chiral superfield $\Phi^\co$, called the chiral compensator,
is primary and has the weights $(\Delta,w)=(1,2/3)$.
The superpotential $W(\Phi)$ and
the gauge holomorphic function $H_{(a)(b)}(\Phi)$ are
holomorphic functions of matter superfields $\Phi^i$, and
hence primary chiral superfields with vanishing weights
$(\Delta,w)=(0,0)$. The indices of $H_{(a)(b)}$ are
symmetric under the exchange $(a)\leftrightarrow (b)$.
If we require the gauge invariance under the internal symmetry, 
$X_{(a)}$ should act on the chiral compensator, the superpotential and
the gauge holomorphic function as
\begin{equation}
\begin{split}
& X_{(a)}\Phi^\co=\df{1}{3}F_{(a)}\Phi^\co,
\quad
X_{(a)}\b\Phi^\co=\df{1}{3}\b{F}_{(a)}\b\Phi^\co,
\qquad X_{(a)}W=-F_{(a)}W,\\
& \quad X_{(a)} H_{(b)(c)}=V_{(a)} H_{(b)(c)}=
-{f_{(a)(b)}}^{(d)}H_{(d)(c)}-{f_{(a)(c)}}^{(d)}H_{(b)(d)}. 
\end{split}
\label{eq: XaPhic}
\end{equation}

\section{Gauge fixing to Poincar\'e superspace}
\label{sec:GF-G}

We first discuss the superconformal 
gauge-fixing condition for going down to the Poincar\'e superspace. 
For the case of non-vanishing superpotential, the chiral 
compensator $\Phi^\co$ can be redefined as
\begin{equation}
\Phi^\co \ \rightarrow\ \Phi^0 = \Phi^\co W^{1/3}.
\label{eq: redef}
\end{equation}
The new chiral compensator $\Phi^0$ still has the weights
$(\Delta,w)=(1,2/3)$. The integrands of matter action become
$\Phi^\co\b\Phi^\co e^{-K/3}=\Phi^0\b\Phi^0e^{-G/3}$ and
$(\Phi^\co)^3W=(\Phi^0)^3$ with 
\begin{equation}
G= K +\ln |W|^2.
\label{eq:GKW}
\end{equation}
Note that this redefinition is possible only when $W\neq0$.
The set of superconformal gauge-fixing
conditions which realizes the canonically normalized EH and RS terms 
and also gives a real gravitino mass is 
\begin{equation}
D,\ A \ \hbox{gauge}:\ \Phi^0=e^{G/6}, \qquad\quad
K_A \ \hbox{gauge}:\ B_M=0.
\label{eq:GF-G}
\end{equation}
One of the virtues of using $\Phi^0$ and $G$ 
is that they are invariant under the gauged internal symmetry:
$X_{(a)}\Phi^0 = X_{(a)}G = 0$, which
follow from Eqs.~(\ref{eq: FJ}) and  (\ref{eq: XaPhic}). 
This invariance property of $\Phi^0$ and $G$ makes it simpler to fix the
superconformal gauge symmetry independently of the internal one. 
So we can simply extend the previous results of the system without 
YM~\cite{bib: KYY}.

The gauge condition $B_M=0$ and the curvature constraint 
$R(D)_{\ul{\alpha}B}=0$ constrain the $K_A$-gauge superfields,
similarly to the case without YM~\cite{bib: B1}.
The following is the restricted form of $K_A$-gauge superfields 
which is needed for later discussion:
\begin{equation}
 f_{\alpha\beta}=-f_{\beta\alpha}=-\epsilon_{\alpha\beta}\b{R},
\quad
f_{\d\alpha\d\beta}=-f_{\d\beta\d\alpha}=\epsilon_{\d\alpha\d\beta} R,
\quad
f_{\alpha\d\beta}=-f_{\d\beta\alpha}
=-\df{1}{2}G_{\alpha\d\beta},
\quad
f_{\ul\alpha b}=-f_{b \ul\alpha}.
\label{eq:SKgauge}
\end{equation}

The equations $\b\na^{\d\alpha} \Phi^0=0$ and 
$\b\na^{\d\alpha}\na_\beta\Phi^0 =-2i \na^{\d\alpha}{}_\beta\Phi^0$,
which follow from the chirality condition of $\Phi^0$, 
determine the $A$-gauge fields in the same 
form as before [(4.12), (4.13) and (4.21) in Ref.\cite{bib: KYY}]. 
Under the gauge-fixing condition (\ref{eq:GF-G}), we obtain
\begin{equation}
\begin{split}
&A_{\alpha} \,=\, \df{i}{4}G_j \cal{D}^{\rm P}_{\,\alpha}{\Phi}^{j},
\qquad
A_{\d\alpha}
= -\df{i}{4}G_{j^*}\b{\cal{D}}^{\rm P}_{\,\d\alpha}\b{\Phi}^{j^*}, \\
&A_\alpha{}^{\d\beta} =
\df{i}{4}(G_i \nabla_\alpha{}^{\d\beta} \Phi^i
-G_{i^*}\nabla_\alpha{}^{\d\beta}
\b\Phi^{i^*})+\df{1}{4}G_{ij^*}\nabla_\alpha\Phi^i
\b\nabla^{\d\beta}\b\Phi^{j^*}-\df{3}{2}{G_\alpha}^{\d\beta},
\label{eq:Agf} 
\end{split}
\end{equation}
where $\cal{D}^{\rm P}_A$ is the covariant derivative in the 
Poincar\'e SUGRA including the YM part and given by
\begin{eqnarray}
\cal{D}^{\rm P}_A &=&
E_A{}^M\partial_M  -\df{1}{2}\phi_A{}^{bc}M_{cb}
-\cal{A}_A^{(a)}X_{(a)} \nn
&=& \nabla_A + A_A \,A
+ B_A D + f_A{}^B K_B.
\label{eq:PoincareDer}
\end{eqnarray}
We further obtain the following expression
for the components of chiral compensator $\Phi^0$ [(4.18) and (4.19) in
Ref.~\cite{bib: KYY}]:
\begin{align}
\left.\na_\alpha\Phi^0\right| &=
\df{1}{3}e^{G/6}G_i \nabla_\alpha\Phi^i\Big|, 
\label{eq: fix_spinor} \\
\left.\na^2\Phi^0\right| &= \df13 e^{G/6} \Bigl(
G_i\na^2\Phi^i +\bigl(G_{ij}+\df{1}{3}G_iG_j\bigr)
\nabla^\alpha\Phi^j \nabla_\alpha\Phi^i -24\b{R} \Bigr)\Bigr|.
\label{eq:hcompensator}
\end{align}
Here the vertical bar ``$|$'' means the $\theta=\b\theta=0$
projection, i.e., the lowest component of superfield. 

The gauge fixing (\ref{eq:GF-G}) is the most convenient and 
physical one. But the redefinition \eqref{eq:GKW} cannot be done
when the system has no superpotential term.
In this case, the most physical gauge fixing is given by 
\begin{equation}
\Phi^\co = e^{K/6}, \qquad B_M=0,
\label{eq:GF-K}
\end{equation}
which also realizes the canonically normalized EH term (if $W\neq0$,
(\ref{eq:GF-K}) generally leads to a complex gravitino mass). 
The main difference is that the gauge fixing \eqref{eq:GF-K} no longer
preserves the internal gauge symmetry, 
since $e^{K/6}$ and $\Phi^\co$ are transformed differently under $X_{(a)}$.
This violation of internal symmetry can be compensated by the $A$-gauge 
transformation. Noting that 
$A \Phi^\co = i\frac{2}{3}\Phi^\co$ and $e^{K/6}$ is 
$A$-gauge invariant, we find that 
the following combination of the internal gauge and $A$ transformations 
rotates $\Phi^\co$ and $e^{K/6}$ in the same way:
\begin{equation}
\tilde{X}_{(a)} = X_{(a)}+\df{i}{4}(F_{(a)}-\b{F}_{(a)})A.
\end{equation}
Therefore $\tilde{X}_{(a)}$ gives the remaining internal gauge symmetry 
after the gauge fixing \eqref{eq:GF-K}, and satisfies the 
same commutation relation as $X_{(a)}$
\begin{equation}
 [\tilde{X}_{(a)},\tilde{X}_{(b)}]=-{f_{(a)(b)}}^{(c)}\tilde{X}_{(c)}.
\end{equation}
This is guaranteed by the relation 
$ V^i_{(a)}{\partial F_{(b)}}/\partial{\Phi^i}
-V^i_{(b)}{\partial F_{(a)}}/\partial{\Phi^i}=-{f_{(a)(b)}}^{(c)}F_{(c)}$,
which is obtained by acting both sides of Eq.~\eqref{eq:Xacom} on
$\Phi^\co$. The $A$ gauge field $A_M$ is no longer inert under 
the internal gauge transformation $\tilde{X}_{(a)}$ but is transformed as
\begin{equation}
 \delta_G(\xi^{(a)}\tilde{X}_{(a)}) A_M 
= \der_M \(\xi^{(a)}\df{i}{4}(F_{(a)}-\b{F}_{(a)})\),
\end{equation}
which matches to the isometric K\"ahler transformation
discussed in~\cite{bib: BG2}. That is, from the superconformal
viewpoint, the isometric K\"ahler transformation is understood to be
the combination of internal gauge and $A$ transformations which leaves
the gauge-fixing condition (\ref{eq:GF-K}) inert.

The covariant derivatives $\tilde{\cal{D}}^{\rm P}_A$ in the  
Poincar\'e SUGRA after imposing (\ref{eq:GF-K}) should be 
defined by using $\tilde{X}_{(a)}$ as
\begin{eqnarray}
\tilde{\cal{D}}^{\rm P}_A &=&
 E_A{}^M\partial_M - \df12 \phi_A{}^{cb}M_{bc}
-\cal{A}_A^{(a)}\tilde{X}_{(a)} \\
&=& \nabla_A + \(A_A 
-\df{i}{4}(F_{(a)}-\b{F}_{(a)})\cal{A}_A^{(a)}\) A 
 + B_A D + f_A{}^B K_B.
\label{eq:Pcd-K}
\end{eqnarray}
The $A$-gauge superfields and the components of chiral compensator 
can be obtained similarly to Eqs.~(\ref{eq:Agf}), (\ref{eq: fix_spinor}) 
and (\ref{eq:hcompensator}) as
\begin{equation}
\begin{split}
&A_{\alpha}= 
\df{i}{4}\(K_j\tilde{\cal{D}}^{\rm P}_{\,\alpha}{\Phi}^{j}
+(F_{(a)}-\b{F}_{(a)})\cal{A}_\alpha^{(a)}\),
\qquad
A_{\d\alpha}= 
\df{i}{4}\(-K_{j^*}\b{\tilde{\cal{D}}}^{\rm P}_{\,\d\alpha}\b{\Phi}^{j^*}
+(F_{(a)}-\b{F}_{(a)})\cal{A}_{\d{\alpha}}^{(a)}\), \\
&A_\alpha{}^{\d\beta}=
 \df{i}{4}\(K_i \tilde{\cal{D}}^{\rm P}_{\,\alpha}{}^{\d\beta} \Phi^i
 -K_{i^*}\tilde{\cal{D}}^{\rm P}_{\,\alpha}{}^{\d\beta}
 \b\Phi^{i^*}
 +(F_{(a)}-\b{F}_{(a)})\cal{A}_{\ \alpha}^{(a)\,\d{\beta}}  \) 
 +\df{1}{4}K_{ij^*}(\tilde{\cal{D}}^{\rm P}_{\,\alpha}\Phi^i)
 (\b{\tilde{\cal{D}}}^{\rm P}{}^{\d\beta}
 \b\Phi^{j^*})-\df{3}{2}{G_\alpha}^{\d\beta},  \\
&  \na_\alpha\Phi^0 \bigr| =
\df{1}{3}e^{K/6}K_i \nabla_\alpha\Phi^i\bigr|, 
\quad
\na^2\Phi^0\bigr| = \df13 e^{K/6} \Bigl(
K_i\na^2\Phi^i +\bigl(K_{ij}+\df{1}{3}K_iK_j\bigr)
\nabla^\alpha\Phi^j \nabla_\alpha\Phi^i -24\b{R} \Bigr)\Bigr|.
\label{eq:GF-K2}
\end{split}
\end{equation}
The expressions for the $A$-gauge superfields are simply obtained from 
(\ref{eq:Agf}) by the replacement $A_A \rightarrow A_A 
-\frac{i}{4}(F_{(a)}-\b{F}_{(a)})\cal{A}_A^{(a)}$ as understood by
the comparison of covariant derivatives.
The gauge-fixed $A$-gauge superfields in (\ref{eq:GF-K2})
exactly agree with the composite gauge potential
in the isometric K\"ahler superspace~\cite{bib: BG2}.

After the gauge fixing, we have the Poincar\'e SUGRA in superspace,
which is the so-called isometric K\"ahler superspace.
For the condition $\Phi^0=e^{G/6}$, the gauge-fixed superspace action becomes
\begin{equation}
 S=\int d^4x d^4\theta\, E\, 
\(-\df{3}{2}
+\df{e^{G/2}}{2R}-\df{1}{8R}H_{(a)(b)}
  \cal{W}^{(a)\alpha}{\cal{W}^{(b)}}_\alpha \)
+\text{h.c.} \ .
\label{eq:action_P}
\end{equation}
Here we have used a relation of F and D-type actions
\begin{equation}
 \int d^4x d^2\theta \cal{E} U = \int d^4x d^4\theta E
\df{U (\Phi^\co\b\Phi^\co e^{-K/3})}
{-\tfrac{1}{4}\b\na^2(\Phi^\co\b\Phi^\co e^{-K/3})},
\end{equation}
where $U$ is a chiral primary superfield with the weights
$(\Delta,w)=(3,2)$. Further, noticing that 
the gauge conditions \eqref{eq:GF-G} and their result 
for the $K_A$-gauge fields \eqref{eq:SKgauge} lead to
\begin{equation}
 -\frac{1}{4}\b\na^2(\Phi^\co\b\Phi^\co e^{-K/3})
=f_{\d\alpha\d\beta}\epsilon^{\d\beta\d\alpha}
(\Phi^\co\b\Phi^\co e^{-K/3}) = 2R,
\end{equation}
one finds that the action in conformal superspace (\ref{eq:action}) 
is reduced to (\ref{eq:action_P}).
For the other gauge fixing $\Phi^\co=e^{K/6}$, 
the resultant superspace action of Poincar\'e SUGRA is similarly
obtained (by an apparent replacement $G\to K+\ln|W|^2$ in 
(\ref{eq:action_P})).

\bigskip

\section{Gauge fixing to component approach}
\label{sec: sstoc}

Next we show the explicit form of superspace gauge fixing
for going down to the superconformal tensor calculus in 
component approach. In conformal superspace, the gauge fields
$h_M{}^{\cal A}$ and the gauge transformation parameters 
$\xi^{\cal A}$ are superfields with higher $\theta$ components, 
and further the gauge superfields contain the spinor-indexed components 
$h_\mu{}^{\cal A}$ and $h^{\dot{\mu}{\cal A}}$ (which we call the
spinor gauge superfields henceforth). 
But there is no such extra freedom of gauge fields in component approach. 
We show in the following that all these extra fields 
can be gauge-fixed to be zero by using extra gauge freedom or 
otherwise be reduced to the known quantities in component approach.

Let us first remark that the following quantities in superspace 
have their counterparts in component approach so that they
are known quantities: 
\begin{enumerate}
\item The lowest component of curved vector gauge superfield,
$h_m{}^{\cal{A}}|$.
This corresponds to $h_\mu{}^A$ in component approach.
\item The lowest components of flat spinor-spinor and spinor-vector 
curvatures, $R_{\ul\alpha\ul\beta}{}^{\cal{A}}|$ and 
$R_{a \ul\beta}{}^{\cal{A}}|$. These correspond to the coefficients 
of the terms appearing on the RHSs of the commutators 
$[\delta_Q,\delta_Q]$ and $[\delta_{\tilde{P}},\delta_Q]$ in component
approach.
\item The lowest components of the covariant derivatives of 
flat spinor-spinor and vector-spinor curvatures
$\na_{\ul\beta} \cdots\na_{\ul\delta}
R_{\ul\alpha\ul\beta}{}^{\cal{A}}|$ and 
$\na_{\ul\beta} \cdots\na_{\ul\delta}R_{a \ul\beta}{}^{\cal{A}}| $.
These correspond to the supersymmetry transformations of the
coefficients in $[\delta_Q,\delta_Q]$ and $[\delta_{\tilde{P}},\delta_Q]$
in component approach.
\end{enumerate}
We show that all other components in the $\theta$ 
expansions of spinor and vector gauge
superfields can be gauge-fixed to zero or be written in terms of 
these known quantities.

\subsection{Gauge conditions for spinor gauge superfields}

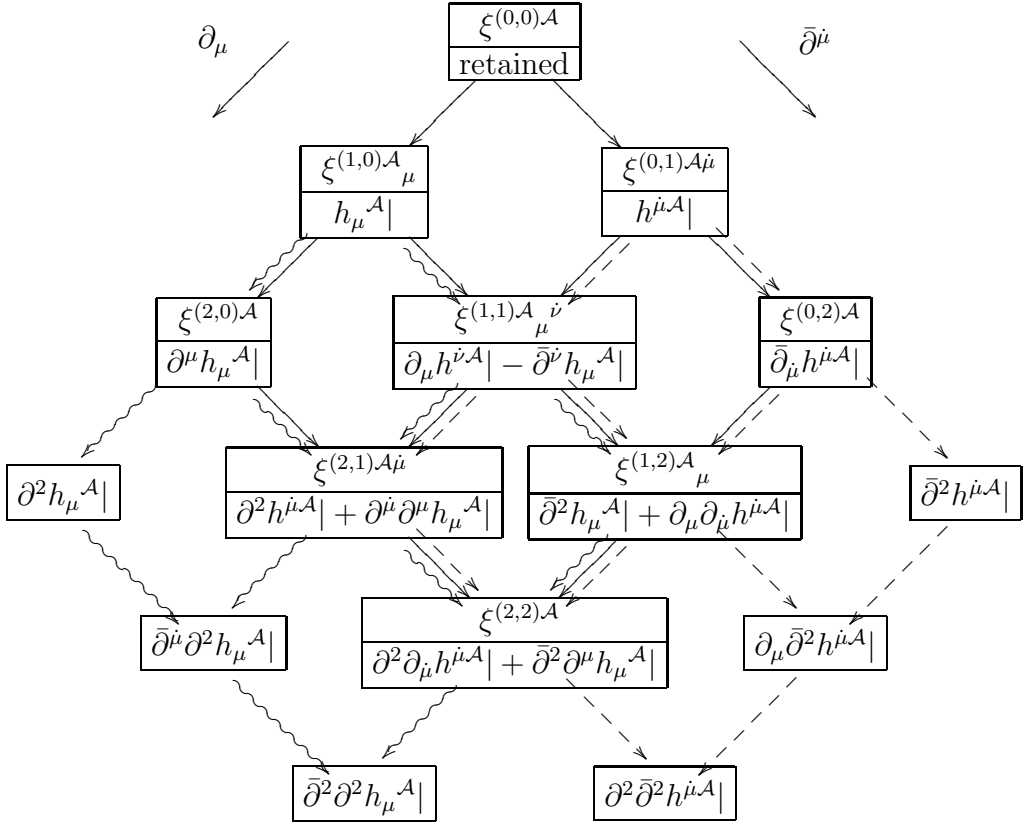
\begin{figure}[t]
\[
\begin{xy}
 (60,120) 
*[F]\txt{
$\Tspan{\xi^{(0,0)\cal{A}}}$
\\ \hline
$\ \Tspan{\text{retained}}\ $
}="00",
 (40,100) 
*[F]\txt{
$\ \Tspan{\xi^{(1,0)\cal{A}}{}_\mu }\ $
\\ \hline
$\ \Tspan{h_\mu{}^{\cal{A}}| }\ $
}="10",
 (80,100) 
*[F]\txt{
$\ \Tspan{ \xi^{(0,1)\cal{A} \d\mu}}\ $
\\ \hline
$\ \Tspan{ h^{\d\mu\cal{A}}|} \ $
}="01",
 (20,80) 
*[F]\txt{
$ \ \Tspan{ \xi^{(2,0)\cal{A}}}\ $
\\ \hline
$\ \Tspan{\der^\mu h_\mu{}^{\cal{A}}|}\ $
}="20",
 (60,80) 
*[F]\txt{
$\ \Tspan{\rule{0pt}{2.1ex} \xi^{(1,1)\cal{A}}{}_{\mu}{}^{\d\nu}}\rule{1em}{0pt} $
\\ \hline
$\ \Tspan{\der_\mu h^{\d\nu\cal{A}}|-\b\der^{\d\nu}h_\mu{}^{\cal{A}}|}\ $
}="11",
 (100,80) 
*[F]\txt{
$\ \Tspan{ \xi^{(0,2)\cal{A}}}\ $
\\ \hline
$\ \Tspan{\b\der_{\d\mu} h^{\d\mu\cal{A}}|}\ $
}="02",
 (0,60) 
*[F]\txt{
\Tspan{\rule{0pt}{2.6ex}\der^2 h_\mu{}^{\cal{A}}|\ }
}="h20",
 (40,60) 
*[F]\txt{
$\ \Tspan{\rule{0pt}{2.1ex} \xi^{(2,1) \cal{A}\d\mu}}\rule{2ex}{0pt}$
\\ \hline
$\ \Tspan{\der^2 h^{\d\mu\cal{A}}|
+\der^{\d\mu}\der^\mu h_{\mu}{}^{\cal{A}}|}\ $
}="21",
 (80,60) 
*[F]\txt{
$\Tspan{\rule{0pt}{2.1ex}\xi^{(1,2) \cal{A}}{}_\mu}\rule{1ex}{0pt}$
\\ \hline
$\Tspan{\ \b\der^2h_{\mu}{}^{\cal{A}}|
+\der_\mu\der_{\d\mu}h^{\d\mu\cal{A}}| \ }$
}="12",
 (120,60) 
*[F]\txt{
\Tspan{\rule{0pt}{2.6ex}\b\der^2h^{\d\mu\cal{A}}|\ }
}="bh02",
 (20,40) 
*[F]\txt{
\Tspan{\rule{0pt}{2.6ex}\b\der^{\d\mu}\der^2h_\mu{}^{\cal{A}}|\ }
}="h21",
 (60,40) 
*[F]\txt{
$\ \Tspan{\xi^{(2,2)\cal{A}}}\ $
\\ \hline
$\ \Tspan{\der^2\der_{\d\mu}h^{\d\mu\cal{A}}|
+\b\der^2\der^\mu h_\mu{}^{\cal{A}}|}\ $
}="22",
 (100,40) 
*[F]\txt{
\Tspan{\rule{0pt}{2.6ex}\der_\mu\b\der^2h^{\d\mu\cal{A}}|\ }
}="bh12",
 (40,20) 
*[F]\txt{
\Tspan{\rule{0pt}{2.6ex}\b\der^2 \der^2h_\mu{}^{\cal{A}}|\ }
}="h22",
 (80,20) 
*[F]\txt{
\Tspan{\rule{0pt}{2.6ex}\der^2 \b\der^2 h^{\d\mu\cal{A}}|\ }
}="bh22",
 (20,120) 
*{\der_\mu}
="der",
 (100,120) 
*{\b\der^{\d\mu}}
="bder",
"00"+(-2,0)="00l",
"01"+(-2,0)="01l",
\ar "00" ; "10"
\ar "00" ; "01"
\ar "10" ; "20"
\ar@<-1.5mm> @{~>} "10" ; "20"
\ar@{->} "10" ; "11"
\ar@<-1.5mm>@{~>} "10" ; "11"
\ar@<1.5mm>@{-->} "01" ; "11"
\ar@{->} "01" ; "11"
\ar@<1.5mm>@{-->} "01" ; "02"
\ar@{->} "01" ; "02"
\ar@<-1.5mm>@{~>} "20" ; "h20"
\ar@{->} "20" ; "21"
\ar@<-1.5mm>@{~>} "20" ; "21"
\ar@{->} "11" ; "21"
\ar @<1.5mm>@{-->} "11" ; "21"
\ar@<-1.5mm>@{~>} "11" ; "21"
\ar@<1.5mm>@{-->} "11" ; "12"
\ar@<-1.5mm>@{~>} "11" ; "12"
\ar@{->} "11" ; "12"
\ar@{->} "02" ; "12"
\ar@<1.5mm>@{-->} "02" ; "12"
\ar@<1.5mm>@{-->} "02" ; "bh02"
\ar@<-1.5mm>@{~>} "h20" ; "h21"
\ar@<-1.5mm>@{~>}"21" ; "h21"
\ar@<-1.5mm>@{~>} "21" ; "22"
\ar@<1.5mm>@{-->} "21" ; "22"
\ar@{->} "21" ; "22"
\ar@<1.5mm>@{-->} "12" ; "22"
\ar@<-1.5mm>@{~>} "12" ; "22"
\ar@{->} "12" ; "22"
\ar@<1.5mm>@{-->} "12" ; "bh12"
\ar@<1.5mm>@{-->} "bh02" ; "bh12"
\ar@<-1.5mm>@{~>} "h21" ; "h22"
\ar@<-1.5mm>@{~>} "22" ; "h22"
\ar@<1.5mm>@{-->} "22" ; "bh22"
\ar@<1.5mm>@{-->} "bh12" ; "bh22"
\ar (30,120) ; (20,110)
\ar (90,120) ; (100,110)
\end{xy}
\]
\caption{{\footnotesize
The $3\times3$ `diamond' (denoted by rigid arrows) 
for the $\theta^n\bar\theta^m$ components 
$\xi^{(n,m)\cal{A}}$ ($n,m=0,1,2$) of the gauge parameter superfield 
$\xi^{\cal{A}}$. The other 
two `diamonds' (denoted by wavy and dashed arrows, respectively) 
for the $\theta^n\bar\theta^m$ components of the 
spinor gauge superfields $h_\mu{}^{\cal{A}}$ and 
$h^{\d\mu\cal{A}}$ are overlapped on it such that the gauge parameter 
$\xi^{(n,m)\cal{A}}$ at each point can be used to fix the spinor gauge field 
components on the same point. The explicit gauge field components which are 
gauge-fixed to zero (or constant) are shown at downstairs of the two-story 
boxes. The gauge field components lying outside of the gauge 
parameter diamond cannot be gauge-fixed but are determined 
by the curvature constraints as shown in the text.}}
\label{fig:sgf}
\end{figure}

The gauge transformation law of spinor gauge superfields is given by
\begin{equation}
 \delta_G(\xi^{\cal{B}}X_{\cal{B}})h_{\ul{\mu}}{}^{\cal{A}}
=\der_{\ul{\mu}}\xi^{\cal{A}}
+h_{\ul{\mu}}{}^{\cal{C}}\xi^{\cal{B}}f_{\cal{BC}}{}^{\cal{A}}.
\label{eq:gaugetrf}
\end{equation}
The gauge-fixing procedure can be seen explicitly as follows.
We first expand the gauge transformation parameter as
\begin{equation}
\begin{split}
 \xi^{\cal{A}}(x,\theta,\b\theta)
&=\xi^{(0,0)\cal{A}}(x)+\theta^{\mu}\xi^{(1,0)\cal{A}}{}_{\mu}(x)
+\b\theta_{\d\mu} \b\xi^{(0,1)\cal{A} \d\mu}(x)  \\
&\quad
+\theta^2\xi^{(2,0)\cal{A}}(x)
+\b\theta^2\b\xi^{(0,2)\cal{A}}(x)
+\theta^{\mu}\b\theta_{\d\mu}\xi^{(1,1)\cal{A}\d\mu}{}_{\mu}(x)  \\
&\quad
+\theta^2\b\theta^{\mu}\xi^{(2,1)\cal{A}}{}_\mu(x)
+\b\theta^2\theta^\mu\xi^{(1,2)\cal{A}}{}_\mu(x)
+\b\theta^2\theta^2\xi^{(2,2)\cal{A}}(x).
\end{split}
\end{equation}
The lowest component $\xi^{(0,0)\cal{A}}(x)$ is 
identified with the usual gauge transformation parameter 
appearing in component approach. We use all the other
higher $\theta$ transformation parameters 
$\xi^{(n,m)\cal{A}}(x)$ ($n+m\ge1$) to fix the components of spinor gauge 
superfields:
\begin{equation}
\label{eq:sgf}
\begin{array}{c|c|c}
\hline\hline
  &\text{parameters}& \text{conditions on spinor gauge superfields}
\\ \hline
\hbox{1st order}& \xi^{(1,0)\cal{A}}{}_{\mu} & 
\Tspan{E_\mu{}^A|=\delta_\mu{}^\alpha,\qquad h_\mu{}^{\cal{A}'}|=0}
\\ \cline{2-3}
&\xi^{(0,1)\cal{A} \d\mu} &
\Tspan{E^{\d\mu A}|=\delta^{\d\mu}{}_{\d\alpha},\qquad h^{\d\mu\cal{A}'}|=0}
\\ \hline
 &\xi^{(2,0)\cal{A}}&\Tspan{\der^\mu h_{\mu}{}^{\cal{A}}|}=0
\\ \cline{2-3}
\hbox{2nd order} & \xi^{(0,2)\cal{A}}&\Tspan{\b\der_{\d\mu} h^{\d\mu\cal{A}}|}=0
\\ \cline{2-3}
&\xi^{(1,1)\cal{A}\d\mu}{}_{\mu}
&\Tspan{\der_{\mu}h^{\d\mu\cal{A}}|
-\b\der^{\d\mu}h_{\mu}{}^{\cal{A}}|=0}
\\ \hline
\hbox{3rd order}& \xi^{(2,1)\cal{A}\d\mu}&
\Tspan{\der^2 h^{\d\mu\cal{A}}|
+\b\der^{\d\mu}\der^\mu h_{\mu}{}^{\cal{A}}| =0}
\\ \cline{2-3}
&\xi^{(1,2)\cal{A}}{}_\mu 
&\Tspan{\b\der^2h_{\mu}{}^{\cal{A}}|+\der_\mu\b\der_{\d\mu}h^{\d\mu 
\cal{A}}|=0}
\\ \hline
\hbox{4th order} &\xi^{(2,2)\cal{A}}
&\Tspan{\der^2\b\der_{\d\mu}h^{\d\mu\cal{A}}|
+\b\der^2\der^\mu h_\mu{}^{\cal{A}}|=0}
\\ \hline
\end{array}
\end{equation}
This set of gauge-fixing conditions is visualized in Fig.~\ref{fig:sgf}.
Note that the fixed gauge field components 
are indeed gauge-variant quantities which are shifted by the 
inhomogeneous transformation 
part $\delta h_\mu^{\cal{A}}=\partial_\mu\xi^{\cal{A}}$. 
For instance, using $\{\partial_\mu, \b\partial^{\d\mu}\}=0$, we have
\begin{equation}
-\varepsilon_{\mu\nu}\der^\rho h_{\rho}{}^{\cal{A}}=
\partial_\mu h_\nu{}^{\cal{A}}-\partial_\nu h_\mu{}^{\cal{A}}, \qquad 
\der^2 h^{\d\mu\cal{A}}
+\b\der^{\d\mu}\der^\mu h_{\mu}{}^{\cal{A}}
=\partial^\mu\( \partial_\mu h^{\d\mu\cal{A}}- \b\der^{\d\mu}h_{\mu}{}^{\cal{A}}\).
\end{equation}
We use the word ``gauge-variant'' when a quantity receives 
an inhomogeneous shift under the gauge transformations with 
$\xi^{(n,m)}(x)$ $(n+m\geq1)$, and otherwise call ``gauge-invariant''. 
The point is that the gauge-fixed quantities in the above table
exhaust the gauge variants. Therefore, once they are fixed
to be zero (or constants), all the other quantities in the spinor
gauge superfields $h_{\mu}{}^{\cal{A}}$ and 
$h^{\d\mu\cal{A}}$ can be expressed by gauge-invariant quantities, 
namely, by covariant curvatures.  

Let us see this more explicitly. 
First we express all the curved spinor derivatives of the curved spinor 
gauge fields, 
$\partial_{\ul{\nu}}\cdots\partial_{\ul{\rho}}h_{\ul{\mu}}{}^{\cal{A}}$,
in terms of the symmetric derivative 
\begin{equation}
\partial_{\ul{\mu}}h_{\ul{\nu}}{}^{\cal{A}}+
\partial_{\ul{\nu}}h_{\ul{\mu}}{}^{\cal{A}},
\end{equation}
which is later rewritten to the covariant curvature with flat spinor
indices. 

At the first order derivative level, 
setting the antisymmetric part equal to zero by the 
above gauge-fixing condition, we find
\begin{equation}
 \der_\nu h_\mu{}^{\cal{A}}|
=\df{1}{2}(\der_\nu h_\mu{}^{\cal{A}}+\der_\mu h_\nu{}^{\cal{A}})| ,
\qquad
 \b\der^{\d\nu} h^{\d\mu\cal{A}}|
=\df{1}{2}(\b\der^{\d\nu} h^{\d\mu\cal{A}}
+\b\der^{\d\mu} h^{\d\nu\cal{A}})|,
\end{equation}
\begin{equation}
\b\der^{\d\nu} h_\mu{}^{\cal{A}}| =
\df{1}{2}(\b\der^{\d\nu} h_\mu{}^{\cal{A}}
+\der_\mu h^{\d\nu\cal{A}})| ,
\qquad
\der_\nu h^{\d\mu\cal{A}}|
=\df{1}{2}(\der_\nu h^{\d\mu\cal{A}}
+\b\der^{\d\mu} h_\nu{}^{ \cal{A}})|.
\end{equation}
At the second order derivative level, the identity
$\der_\mu\der^\nu h_{\nu}= \frac{1}{2}\der^2 h_\mu$ leads to
\begin{equation}
 \der^2 h_\mu{}^{\cal{A}}| =
\df{2}{3}\der^\nu(\der_\nu h_\mu{}^{\cal{A}} 
+ \der_\mu h_\nu{}^{\cal{A}})|,
\qquad
\b\der^2 h^{\d\mu\cal{A}}| =
\df{2}{3}\b\der_{\d\nu}(\b\der^{\d\nu} h^{\d\mu\cal{A}} 
+ \b\der^{\d\mu} h^{\d\nu\cal{A}})|,
\label{eq:sgsym2}
\end{equation}
which actually hold as the superfield equations without the vertical
bars `$|$'. 
If we use the above gauge conditions by $\xi^{(2,1) \cal{A} \d\mu}$ and 
$\xi^{(1,2)\cal{A}}{}_\mu$ we also have
\begin{equation}
\begin{split}
 \b\der^2 h_\mu{}^{\cal A}
&= \frac12 \b\der_{\d\nu}
(\b\der^{\d\nu} h_\mu{}^{\cal A} +
\der_{\mu} h^{\d\nu\cal A} )|,
\qquad
 \der^2 h^{\d\mu\cal A}|
= \frac12\der^{\nu}
(\der_{\nu} h^{\d\mu\cal A} +
\b\der^{\d\mu} h_\nu{}^{ \cal A} )|,  \\
\der_\rho\b\der^{\d\nu} h_\mu{}^{\cal{A}}|
&= \df{1}{2}
\der_\rho(\b\der^{\d\nu} h_\mu{}^{\cal{A}}
+\der_\mu h^{\d\nu\cal{A}})|
-\df{1}{4} \b\der^{\d\nu}
(\der_\rho h_\mu{}^{\cal{A}}+\der_\mu h_\rho{}^{\cal{A}}) |,  \\
\b\der^{\d\rho}\der_{\nu} h^{\d\mu\cal{A}}|
&= \df{1}{2}
\b\der^{\d\rho} (\der_\nu h^{\d\mu\cal{A}}
+\b\der^{\d\mu} h_\nu{}^{ \cal{A}})|
-\df{1}{4} \der_\nu 
(\b\der^{\d\rho} h^{\d\mu\cal{A}}
+\b\der^{\d\mu} h^{\d\rho\cal{A}}) |.
\end{split}
\end{equation}
At the third order derivative level, Eq.~\eqref{eq:sgsym2} without 
the bars leads to 
\begin{equation}
\b\der^{\d\nu}\der^2 h_\mu{}^{\cal{A}}|
= \df{2}{3}
\b\der^{\d\nu}\der^\rho 
(
\der_\rho h_\mu{}^{\cal{A}}
+\der_\mu h_\rho{}^{\cal{A}}
)|,
\qquad
\der_\nu\b\der^2 h^{\d\mu\cal{A}}|
= \df{2}{3}
\der_\nu\b\der_{\d\rho}
(
\b\der^{\d\rho} h^{\d\mu\cal{A}}
+\b\der^{\d\mu} h^{\d\rho\cal{A}}
)|.
\end{equation}
Using the gauge conditions by $\xi^{(2,2)}$, we have
\begin{equation}
\begin{split}
 \der_\nu\b\der^2 h_\mu{}^{\cal{A}}|
&=\df{1}{2}\der_\nu\b\der_{\d\rho}
(\b\der^{\rho}h_\mu{}^{\cal{A}}
+\der_\mu h^{\d\rho\cal{A}}
)|
+\df{1}{4}\b\der^2
(
\der_\nu h_{\mu}{}^{\cal{A}}+
\der_\mu h_{\nu}{}^{\cal{A}}
)|, \\ 
 \b\der^{\d\nu}\der^2 h^{\d\mu\cal{A}}|
&= \df{1}{2}\b\der^{\d\nu}\der^\rho 
(\der_\rho h^{\d\mu\cal{A}}
+\b\der^{\d\mu} h_\rho{}^{ \cal{A}}
)|
+\df{1}{4}\der^2
(
\b\der^{\d\nu} h^{\d\mu\cal{A}}+
\b\der^{\d\mu} h^{\d\nu\cal{A}}
)|.
\end{split}
\end{equation}
Finally at the fourth order level, Eq.~\eqref{eq:sgsym2} leads to 
\begin{equation}
 \b\der^2\der^2 h_\mu{}^{\cal{A}}|
=\df{2}{3}\b\der^2\der^\nu 
(
\der_\nu h_\mu{}^{\cal{A}}
+
\der_\mu h_\nu{}^{\cal{A}}
)|,
\qquad
 \der^2\b\der^2 h^{\d\mu\cal{A}}|
=\df{2}{3}\der^2\b\der_{\d\nu}
(
\b\der^{\d\nu} h^{\d\mu\cal{A}}
+
\b\der^{\d\mu} h^{\d\nu\cal{A}}
)|.
\end{equation}

In this way, all the curved spinor derivatives of the curved spinor gauge
superfields are written in terms of the curved spinor derivatives on 
the symmetrized 
term $\der_{\ul\nu}h_{\ul\mu}{}^{\cal{A}} 
+\der_{\ul\mu}h_{\ul\nu}{}^{\cal{A}}$. 
We show in the following that they are written by the flat-indexed
curvatures $R_{CB}{}^{\cal{A}}$ and their covariant spinor derivatives 
in the present gauge.

Noting the definition of curvatures
\begin{equation}
  {R_{MN}}^{\cal{A}}
=\der_M{h_N}^{\cal{A}}-\der_N{h_M}^{\cal{A}}
-({E_N}^{C}{h_M}^{\cal{B}'}-{E_M}^{C}{h_N}^{\cal{B}'})
{f_{\cal{B}'C}}^{\cal{A}} -
{h_N}^{\cal{C}'}{h_M}^{\cal{B}'}{f_{\cal{B'C'}}}^{\cal{A}}
\label{eq:defcurv}
\end{equation}
and $h_{\ul{\mu}}{}^{\cal{A'}}|=0$, we see that 
$(\der_{\ul\nu}h_{\ul\mu}
{}^{\cal{A}} +\der_{\ul\mu}h_{\ul\nu}{}^{\cal{A}})|$ is written in terms 
of curvatures:
\begin{equation}
(\der_{\ul{\mu}}h_{\ul{\nu}}{}^{\cal{A}}
+\der_{\ul{\nu}}h_{\ul{\mu}}{}^{\cal{A}})| = 
\delta_{\ul\mu}{}^{\ul{\alpha}}\delta_{\ul\nu}{}^{\ul\beta}
R_{\ul{\alpha}\ul{\beta}}{}^{\cal{A}}| ,
\label{eq:sym0}
\end{equation}
which means that 
$(\der_{\ul\nu}h_{\ul\mu}{}^{\cal{A}}
+\der_{\ul\mu}h_{\ul\nu}{}^{\cal{A}})|$
is written in the component language.
Note also that the lowest components of curved indexed curvatures are
always rewritten by flat indexed ones and can be expressed  
in terms of the fields in component approach as
\begin{equation}
 R_{\ul\nu\ul\mu}{}^{\cal{A}}|
=\delta_{\ul\nu}{}^{\ul\gamma} \delta_{\ul\mu}{}^{\ul\beta} 
R_{\ul\gamma\ul\beta} {}^{\cal{A}}|,
\qquad
 R_{\ul\nu m}{}^{\cal{A}}|
=\delta_{\ul\nu}{}^{\ul\gamma}e_m{}^b R_{\ul\gamma b}{}^{\cal{A}}|
-\df{1}{2}\delta_{\ul\nu}{}^{\ul\gamma}\psi_m{}^{\ul\beta}
R_{\ul\gamma\ul\beta}{}^{\cal{A}}|.
\end{equation}
Then the second order derivative 
$\der_{\ul\rho}(\der_{\ul\nu}h_{\ul\mu}{}^{\cal{A}}
+\der_{\ul\mu}h_{\ul\nu}{}^{\cal{A}})|$ is similarly found to be 
written by the curvatures
\begin{equation}
 \der_{\ul\rho}
(\der_{\ul\nu}h_{\ul\mu}{}^{\cal{A}}
+\der_{\ul\mu}h_{\ul\nu}{}^{\cal{A}}
)|
=\der_{\ul\rho}
R_{\ul\nu\ul\mu}{}^{\cal{A}}| +
\df{1}{2}
({\delta_{\ul\mu}}^{\ul\gamma}R_{\ul\rho\ul\nu}{}^{\cal{B}'}
+{\delta_{\ul\nu}}^{\ul\gamma}R_{\ul\rho\ul\mu}{}^{\cal{B}'})
{f_{\cal{B}'\ul\gamma}}^{\cal{A}} |.
\end{equation}
The first term in the r.h.s., the spinor derivative on curved indexed
curvatures $\der_{\ul\rho}R_{\ul\nu\ul\mu}{}^{\cal{A}}|$ is nontrivial,
but written by the covariant derivatives on flat indexed 
curvatures as
\begin{equation}
\begin{split}
 \der_{\ul\rho} R_{\ul\nu\ul\mu}{}^{\cal{A}}|
&= \der_{\ul\rho}\bigl(E_{\ul\nu}{}^CE_{\ul\mu}{}^B 
R_{CB}{}^{\cal{A}}\bigr)\bigr|  \\
&= \df{1}{2}R(P)_{\ul\rho\ul\nu}{}^C
\delta_{\ul\mu}{}^{\ul\beta} 
R_{C\ul\beta}{}^{\cal{A}}| 
-\df{1}{2}
\delta_{\ul\nu}{}^{\ul\gamma} R(P)_{\ul\rho\ul\mu}{}^B 
R_{\ul{\gamma}B}{}^{\cal{A}}| +
\delta_{\ul\nu}{}^{\ul\gamma}\delta_{\ul\mu}{}^{\ul\beta} 
\delta_{\ul\rho}{}^{\ul\delta}\na_{\ul\delta}
R_{\ul\gamma\ul\beta}{}^{\cal{A}}|. 
\end{split}
\label{eq:curv1}
\end{equation}
Here we have used the definition of (the lowest component of) the torsion 
$R(P)_{\rho\nu}{}^{C}|=(\der_\rho E_\nu{}^C+\der_\nu E_\rho{}^C)|$
and the fact that $\der_\mu$ in the present gauge is equal to 
$\delta_{\mu}{}^{\alpha}\na_\alpha$ at the lowest level, that is,
\begin{equation}
 \der_{\ul\mu} \Phi|
=(\der_{\ul\mu}-h_{\ul\mu}{}^{\cal{A}'}X_{\cal{A}'})\Phi|
=\delta_\mu{}^\alpha\na_\alpha\Phi|
\end{equation}
on a covariant quantity $\Phi$. 
Therefore $\der_{\ul\rho}(\der_{\ul\nu}h_{\ul\mu}{}^{\cal{A}}
+\der_{\ul\mu}h_{\ul\nu}{}^{\cal{A}})|$ is written in terms of 
the fields in component approach, and so are all the second order derivative 
of the spinor gauge fields. 

The same is true for the third order derivative 
$\der_{\ul\sigma} \der_{\ul\rho}(\der_{\ul\nu}h_{\ul\mu}{}^{\cal{A}}
+\der_{\ul\mu}h_{\ul\nu}{}^{\cal{A}})|$.
This is also written as 
\begin{equation}
\der_{\ul\sigma} \der_{\ul\rho}
(\der_{\ul\nu}h_{\ul\mu}{}^{\cal{A}}
+\der_{\ul\mu}h_{\ul\nu}{}^{\cal{A}}
)|
=\der_{\ul\sigma}\der_{\ul\rho}
R_{\ul\nu\ul\mu}{}^{\cal{A}}\bigr|
-\der_{\ul\sigma}\der_{\ul\rho}
\bigl(({E_{\ul\mu}}^{C}{h_{\ul\nu}}^{\cal{B}'}
+{E_{\ul\nu}}^{C}{h_{\ul\mu}}^{\cal{B}'})
{f_{\cal{B}'C}}^{\cal{A}} 
+
{h_{\ul\mu}}^{\cal{C}'}{h_{\ul\nu}}^{\cal{B}'}
{f_{\cal{B'C'}}}^{\cal{A}}
\bigr)\bigr|.
\end{equation}
The terms other than $\der_{\ul\sigma}\der_{\ul\rho}
R_{\ul\nu\ul\mu}{}^{\cal{A}}|$ contain at most second order derivatives 
of gauge fields and can be written in component language as shown above.
The non-trivial term $\der_{\ul\sigma}\der_{\ul\rho}
R_{\ul\nu\ul\mu}{}^{\cal{A}}|$ is expanded as
\begin{equation}
 \der_{\ul\sigma}\der_{\ul\rho}
R_{\ul\nu\ul\mu}{}^{\cal{A}}|
=\delta_{\ul\sigma}{}^{\ul\delta}
\delta_{\ul\rho}{}^{\ul\gamma}
\delta_{\ul\nu}{}^{\ul\beta}
\delta_{\ul\mu}{}^{\ul\alpha}
 \na_{\ul\delta}\na_{\ul\gamma}
R_{\ul\beta\ul\alpha}{}^{\cal{A}}|
+\cdots,
\end{equation}
where $\cdots$ denotes the terms which are written by 
the curvatures and their first order covariant spinor derivatives with 
coefficients of at most second order derivatives 
of the gauge fields. Thus the second order spinor derivatives of the 
curved indexed curvatures and hence all the third order derivatives of 
the gauge fields are shown to be written in terms of the known fields in 
component approach. 
The third order curved spinor derivatives on curved indexed curvatures
is similarly given by at most third order flat indexed covariant spinor
derivatives on flat indexed curvatures.
Thus, we have shown all orders of derivative of curved spinor gauge
superfields are written in terms of the known fields in component approach.

The above procedure also makes sense for the gauge fixing in 
Poincar\'e superspace approach to the component SUGRA (the route III
in Fig.~\ref{fig:aim}), since we only use the general definitions of
curvatures and inhomogeneous terms in the gauge transformation laws.
The only difference is the constraints on curvatures, i.e., their
explicit forms after gauge fixing. 

We comment on other possible gauge-fixing conditions.
In the treatment of Ref.~\cite{bib: B1},
the dotted spinor gauge fields are gauge-fixed at superfield
level to be $E^{\d\mu A}=\delta^{\d\mu}{}_{\d\alpha}$ and
$h^{\d\mu\cal{A}'}=0$. In this case, 
$\xi^{(1,1)\cal{A}\d\mu}{}_{\mu}$ is used for realizing
$ \der_{\mu}h^{\d\mu\cal{A}}|=0 $, and
$\xi^{(2,1)\cal{A}\d\mu}$ for $\der^2 h^{\d\mu\cal{A}}|=0$, and so
on. On the other hand, the 
undotted spinor gauge superfields $h_\mu{}^{\cal{A}}$ generally
remain unfixed. Another different example of gauge fixing will be
discussed in section~\ref{sec: corr} for 
the superspace counterpart of the improved gauge in component approach.

\subsection{Higher $\theta$ components of vector gauge superfields}

In the previous section we have shown that all the higher $\theta$
components of spinor gauge superfields are properly gauge-fixed,
and have already used all the $\theta$ components of the gauge 
transformation parameters other than the lowest. 
But there still exist the vector gauge superfields 
to be gauge-fixed for going down to the tensor calculus.
So it is non-trivial whether all the higher $\theta$ components of 
vector gauge superfields are written in terms 
of known fields in component approach.
By using the gauge conditions \eqref{eq:sgf} and the definition of
curvature $R_{\ul\mu n}{}^{\cal{A}}$ in Eq.~\eqref{eq:defcurv}, we
find $\der_{\ul\nu} h_n{}^{\cal{A}}$ is written as
\begin{equation}
\begin{split}
\der_{\ul\mu} h_n{}^\cal{A}|
&=\delta_{\ul\mu}{}^{\ul\delta}E_{n}{}^E R_{\ul\delta E}{}^{\cal{A}}|
-{\delta_{\ul\mu}}^{C}{h_n}^{\cal{B}'}|{f_{\cal{B}'C}}^{\cal{A}}  \\
&=\delta_{\ul\mu}{}^{\ul\delta}e_{n}{}^e R_{\ul\delta e}{}^{\cal{A}}|
+(-)^1\delta_{\ul\mu}{}^{\ul\delta}\tfrac{1}{2}\psi_{n}{}^{\ul\epsilon} 
R_{\ul\delta\ul\epsilon}{}^{\cal{A}}|
-{\delta_{\ul\mu}}^{\ul\gamma}{h_n}^{\cal{B}'}|
{f_{\cal{B}'\ul\gamma}}^{\cal{A}}.
\label{eq:vsexp-1}
\end{split}
\end{equation}
This shows that the first order spinor derivative is expressed in
terms of the fields in component approach. The analysis is performed
in similar ways for the other higher components.
For example, using Eq.~\eqref{eq:defcurv}, we obtain
\begin{equation}
\begin{split}
\der_{\ul\rho}\der_{\ul\mu} h_{n}{}^{\cal{A}}|
&=
\der_{\ul\rho}{R_{{\ul\mu} n}}^{\cal{A}}|
+\der_n\der_{\ul\rho} {h_{\ul\mu}}^{\cal{A}}|  \\
&\quad
+((\der_{\ul\rho}{E_n}^{C}){h_{\ul\mu}}^{\cal{B}'}
+{E_n}^{C}(\der_{\ul\rho}{h_{\ul\mu}}^{\cal{B}'})
-(\der_{\ul\rho}{E_{\ul\mu}}^{C}){h_n}^{\cal{B}'}
-{E_{\ul\mu}}^{C}(\der_{\ul\rho}{h_n}^{\cal{B}'})
)|
{f_{\cal{B}'C}}^{\cal{A}} \\
&\quad
+(\der_{\ul\rho}{h_n}^{\cal{C}'})
{h_{\ul\mu}}^{\cal{B}'}|{f_{\cal{B'C'}}}^{\cal{A}}
+{h_n}^{\cal{C}'}(\der_{\ul\rho}{h_{\ul\mu}}^{\cal{B}'})|
{f_{\cal{B'C'}}}^{\cal{A}}.
\end{split}
\end{equation}
The spinor derivative of curved indexed
curvatures can be replaced by its covariant derivative,
\begin{equation}
\begin{split}
\der_{\ul\mu}{R_{\ul\nu p}}^{\cal{A}}|
&=\der_{\ul\mu}E_{\ul\nu}{}^CE_{p}{}^B{R_{C B}}^{\cal{A}}|\\
&=
(\der_{\ul\mu}E_{\ul\nu}{}^C)E_{p}{}^B{R_{C B}}^{\cal{A}}|
-E_{\ul\nu}{}^C(\der_{\ul\mu}E_{p}{}^B){R_{C B}}^{\cal{A}}|
-E_{\ul\nu}{}^CE_{p}{}^B
\delta_{\ul\mu}{}^{\ul\delta}\na_{\ul\delta}{R_{C B}}^{\cal{A}}|.
\end{split}
\end{equation}
Therefore the second order spinor derivative of vector gauge
superfields are found to be written by the curvatures, their covariant
derivatives, and the lowest components of curved vector superfields. 
For higher $\theta$ terms, we need to evaluate higher order 
spinor derivatives on curvatures which become
\begin{equation}
\begin{split}
\der_{\ul\mu}\der_{\ul\nu}R_{PQ}{}^{\cal{A}}|
&= 
\der_{\ul\mu}\der_{\ul\nu}E_P{}^C E_Q{}^D R_{CD}{}^{\cal{A}}|
=E_P{}^C E_Q{}^D E_{\ul\mu}{}^G E_{\ul\nu}{}^F 
\na_G \na_F R_{CD}{}^{\cal{A}}|
+\cdots ,
\end{split}
\end{equation}
where $\cdots$ means the first order spinor derivatives on curvatures
and/or the second order derivatives on vector gauge superfields.
These have already been shown to be expressed by component approach
fields. The same is clearly true for the third order derivative
of curvatures. 
In the end, under the gauge conditions \eqref{eq:sgf},
all order $\theta$ components in curved vector and spinor 
gauge superfields are written in the language of component approach,
and hence properly gauge-fixed.

\subsection{Higher $\theta$ gauge invariance in component approach} 

We have shown in superspace approach that all the extra fields other
than those appearing in component approach can be eliminated by using
the higher $\theta$ gauge degrees of freedom, 
$\xi^{(n,m){\cal A}}(x)$ $( n+m\geq1 )$. 
We here add an interesting remark which might sound surprising:

{\it All the fields appearing in component approach are in fact 
higher $\theta$ gauge invariant, i.e., invariant under the gauge 
transformation with parameters $\xi^{(n,m){\cal A}}(x)$ 
$( n+m\geq1 )$  but with $\xi^{(0,0){\cal A}}(x)=0$. }

The gauge fields in component approach correspond to the lowest 
components of the gauge superfields with {\it vector} index
$h_m{}^{\cal A}|$, like the vierbein 
$E_m{}^a| = e_m{}^a$ and the gravitino 
$E_m{}^{\ul{\alpha}}| = \psi_m{}^{\ul{\alpha}}$.
Taking the lowest component of the general gauge transformation law 
(\ref{eq:gaugetrf}) for the vector index case ${\ul \mu}\to m$, we have  
\begin{equation}
 \delta_G(\xi^{\cal{B}}X_{\cal{B}})h_m{}^{\cal{A}}|
=\der_m\xi^{\cal{A}}|
+h_{m}{}^{\cal{C}}\,\xi^{\cal{B}}|\,f_{\cal{BC}}{}^{\cal{A}}.
\end{equation}
This contains only the lowest component of gauge parameter
$\xi^{\cal{A}}|=\xi^{(0,0)\cal{A}}(x)$ and its spacial derivative, 
and does not involve any higher $\theta$ gauge parameter
$\xi^{(n,m){\cal A}}(x)$ $(n+m\geq1)$. Therefore the gauge fields 
in component approach are higher $\theta$ gauge invariant. 

Furthermore, {\it all component fields of matter multiplets are also 
higher $\theta$ gauge invariant}, provided that they are identified with 
the lowest components of the {\it covariant derivatives} of matter
superfields. For instance, the component fields in a general 
matter multiplet $[\,\cal{C}, \cal{Z}, \cal{H}, \cdots\,]$ 
in component approach are expressed by using the
covariant derivatives of a primary superfield $\Phi$ as \cite{bib: KYY}
\begin{equation}
\cal{C} = \Phi|, \qquad 
\cal{Z}_\alpha = -i\na_\alpha \Phi|, \qquad 
\cal{H} =
\df{1}{4}(\na^2 \Phi + \b{\na}^2 \Phi)|, 
\quad \cdots .
\end{equation}
Since these covariant derivative quantities are literally covariant, 
their gauge transformation contain no derivative of the gauge parameter 
superfield $\xi^{\cal A}(x,\theta,\b\theta)$ so that their 
lowest components contain only the lowest component 
$\xi^{\cal A}| = \xi^{(0,0){\cal A}}$, e.g.\ for the first derivative,
\begin{equation}
\delta(\xi^AP_A)\nabla_\alpha\Phi| 
= \xi^A| \, \nabla_A (\nabla_\alpha\Phi)|,
\end{equation}
and hence higher $\theta$ gauge invariant.

In this sense, the invariant action formulas in component approach 
are not only invariant under the usual gauge transformation with the
parameter $\xi^{\cal A}(x) = \xi^{(0,0){\cal A}}$, but fully gauge
invariant with the superfield parameter
$\xi^{\cal A}(x,\theta,\bar\theta)$. 
For instance, consider the F-type action formula of conformal
SUGRA~\cite{{Kugo:1982cu},{bib: B1}}
\begin{equation}
 S^{\text{comp}}_F = \int d^4 x\,e
\Bigl(-\df{1}{4}\na^2 W
+\df{i}{2}\b\psi_{a\d\alpha}(\b\sigma^a)^{\d\alpha\beta}\na_\beta W
-\(\b\psi_a\b\sigma^{ab}\b\psi_{b}\)W
\Bigr)\Big|+\text{h.c.}.
\label{eq:Fcomp-action}
\end{equation}
The integrands is written in terms of the vierbein $e_m{}^a$, 
the gravitino $\psi_m{}^{\ul\alpha}$, and the lowest components of the 
covariant superfields, $W|$, $\na_\alpha W|$, $\na^2 W|$.
All of these quantities are higher $\theta$ gauge invariant, and hence
the action is fully gauge invariant with superfield transformation
parameters.

Indeed the above component expression for the F-type action formula
was derived in \cite{bib: B1} starting from the superspace action
\begin{equation}
S^{\text{SS}}_F = \int d^4x d^2\theta\, \cal{E}W +\text{h.c.} ,
\end{equation}
which has the manifest superfield gauge invariance.
While his derivation uses a particular gauge fixing for the higher 
$\theta$ gauge symmetry, the final 
component action $S^{\text{comp}}_F$ (\ref{eq:Fcomp-action}) 
is also superfield gauge invariant. 
Both expressions $S^{\text{comp}}_F$ and $S^{\text{SS}}_F$ are 
fully superfield gauge invariant and coincide with each other in a 
particular gauge, so implying that they coincide in any gauge. 

\bigskip

\section{Correspondence of superconformal gauge fixing}
\label{sec: corr}

In this section, we clarify more explicitly the correspondence of
superconformal gauge fixing between the superspace and component
approaches. In component approach, the gauge-fixing condition that 
leads to the canonically normalized EH and RS terms was firstly given by 
Kugo-Uehara (KU) in Ref.~\cite{bib: KUigc}, called the KU gauge in
what follows.

\subsection{Correspondence of YM sectors}

The following is the translational dictionary for the YM sector in
the YM matter coupled conformal SUGRA
between the superspace formulation in section~2 and
the component one in Ref.~\cite{bib: KUigc}.
\begin{equation}
\begin{array}{c|c}
\hline\hline
\text{component} & \text{superspace}
\\ \hline
\Tspan{B^\alpha_\mu,\ \ F^\alpha_{\mu\nu},\ \ \hat{F}^\alpha_{ab}}
& 
\Tspan{\cal{A}_m^{(a)}|,\ \ \cal{F}^{(a)}_{mn}|,\ \ \cal{F}^{(a)}_{ab}|}
\\ \hline
\Tspan{W^\alpha_{\text{R}},\ \ W^\alpha_{\text{L}}} & 
\Tspan{\cal{W}_\alpha^{(a)}|,\ \ -\cal{W}^{(a)\d\alpha}|}
\\ \hline
D^\alpha,\ \ f_{\alpha\beta}
& \Tspan{\frac{i}{2}\na^\alpha\cal{W}_\alpha^{(a)}|}
=\frac{i}{2}\b\na_{\d\alpha}\cal{W}^{(a)\d\alpha}|,\ \
H_{(a)(b)}|
\\ \hline
i{{T^\alpha}_i}^j z_j,\ \ -iz^{*j}{{T^\alpha}_{j}}^{i}
&
\Tspan{V^i_{(a)}|,\ \ \b{V}^{i^*}_{(a)}}|
\\ \hline
\end{array}
\end{equation}
In this dictionary, we have rescaled the gauge fields and the gaugino
multiplet of internal symmetry by the gauge coupling $\tilde{g}$ such as
$\tilde{g}B^{\alpha}_\mu\to B^{\alpha}_\mu$.
In Ref.~\cite{bib: KUigc}, the internal symmetry was discussed
only for the linear case where the Killing vectors are given by the
representation matrices. 
More general cases of Killing vectors which preserve the K\"ahler
structure were constructed in the framework of superconformal tensor
calculus~\cite{bib: KKLVP}. With the same form of K\"ahler potential, 
our $V^i_{(a)}|$, $\b{V}_{(a)}^{i^*}|$, 
$F_{(a)}|$ and $J_{(a)}|$ correspond to $k_{\alpha i}$, $k^i_\alpha$,
$3r_{\alpha}$ and $\cal{P}_\alpha(z,z^*)$ in Ref.~\cite{bib: KKLVP},
respectively.

\subsection{Non-vanishing superpotential case}

As in the system without YM gauge fields, 
the gauge-fixing conditions in superspace~\eqref{eq:GF-G} 
are found to correspond to the KU gauge:
\begin{equation}
\begin{split}
z_0 = \sr{3} \phi^{-\frac{1}{2}}(z,z^*) = e^{-\cal{G}/6} 
\quad&\leftrightarrow\quad
\Phi^0\big| = e^{G/6}\big| , \\
\chi_{\text{R}0}=-z_0\phi^{-1}\phi^i\chi_{\text{R}i}
=-\df{1}{3}e^{-\cal{G}/6}\cal{G}^i \chi_{\text{R}i}
\quad&\leftrightarrow\quad
\na_\alpha\Phi^0\big| =
\df{1}{3}e^{G/6}G_i \nabla_\alpha\Phi^i\big|,
\label{eq: spi_corr} 
\end{split}
\end{equation}
where $z_0$ and $\chi_{\text{R}0}$ are the lowest and spinor
components of the chiral compensator in the superconformal tensor
calculus, and we have used the correspondence of conformal multiplets,
developed in \cite{bib: KYY}. Unlike Ref.~\cite{bib: KYY}, the
covariant derivative $\na_\alpha$ now contains the gauge field of
internal symmetry.

Among various correspondences, we here focus on the correspondence of
the Poincar\'e supersymmetry after gauge fixing.
Since the spinor derivatives are identified with the supersymmetry 
transformations both in the conformal and Poincar\'e superspaces, 
Eq.~(\ref{eq:PoincareDer})
leads to the Poincar\'e supersymmetry transformation defined by
the superconformal gauge transformations
\begin{equation}
 \delta_G(\eta^{\ul{\alpha}} Q^{\text{P}}_{\ul{\alpha}})
=\delta_G(\eta^{\ul{\alpha}}Q_{\ul{\alpha}})
+\delta_G(\xi(A)'(\eta)A)
+\delta_G(\xi(K)'(\eta)^BK_B)
\label{eq:omQ}
\end{equation}
with the transformation parameters 
$\xi(A)'(\eta)=\eta^\alpha A_{\alpha}+\b\eta_{\d\alpha}A^{\d\alpha}$
and $\xi(K)'(\eta)^A=\eta^\beta f_{\beta}{}^A+
\b\eta_{\d\beta}f^{\d\beta}{}^A$. 
Using the explicit form of the spinor gauge fields 
$A_{\ul\alpha}$ obtained in (\ref{eq:Agf}), we have
the $A$-transformation parameter
\begin{equation}
\xi(A)'(\eta)| \,=\,
\frac{i}4G_j \eta^\alpha\cal{D}_\alpha^{\rm P}\Phi^j\big|
-\frac{i}4G_{j*}\b{\eta}_{\d{\alpha}} \b{\cal{D}}^{{\rm P}\,\d{\alpha}}
\b{\Phi}^{j*}\big|,
\label{eq: xis}
\end{equation}
which is the exactly same
form as the previous result in the system without YM
(Eq.~(4.28) in \cite{bib: KYY}), though the covariant 
derivatives in the above contain the YM covariantization terms. 
The spinor $K_A$-gauge fields $f_{\ul{\alpha}}{}^A$ under the gauge
fixing \eqref{eq:GF-G} are also evaluated by using 
(\ref{eq:SKgauge}) and (\ref{eq:GF-K2}), 
and the resultant transformation parameters are given by
\begin{equation}
\begin{split}
\xi(K)'(\eta)_\alpha\big|
& = -\df{1}{8} \Bigl( e^{-G/6}\nabla^2 \Phi^0
 -\df{1}{3}G_i \na^2\Phi^i\Bigr)\eta_\alpha \big|  \\
&\hspace*{15mm}
-\df{1}{12}
\Big(\Bigl(
    G_{ij}+\df{1}{3}G_iG_j
  \Bigr)
  \eta^\gamma
    \nabla_\gamma\Phi^j
  +G_{ij^*}
    \b\eta_{\d\beta}
    \b\na^{\d\beta}\b\Phi^{j^*}
\Big)
\na_\alpha\Phi^i|  \\
&\hspace*{15mm}
-\df{1}{12}{e_c}^m
\(
  G_i\nabla_m \Phi^i 
  -G_{i^*}\nabla_m\b\Phi^{i^*}
\)
i\sigma^c_{\alpha\d\beta}\b\eta^{\d\beta}\big|
-\frac{i}{3}{e_c}^m A_m
i\sigma^c_{\alpha\d\beta}\b\eta^{\d\beta}\big|,
\end{split}
\label{eq:PQS}
\end{equation}
\begin{equation}
\begin{split}
\xi(K)'(\eta)_a \big|
&= -e_a{}^m ({f_m}^\beta \eta_\beta +f_{m\d\beta}\b\eta^{\d\beta})  \\
& \hspace*{15mm} +\frac{1}{2}e_a{}^m\psi_m{}^\alpha 
\big(\eta^\beta f_{\beta\alpha}
+\b\eta_{\d\beta}{f^{\d\beta}}_\alpha\big)\big|
+\frac{1}{2}e_a{}^m\b\psi_{m\d\alpha}
\big(\eta^\beta{f_\beta}^{\d\alpha}
+\b\eta_{\d\beta}{f^{\d\beta\d\alpha}}\big)
\big| \,.
\end{split}
\label{eq:PQK} 
\end{equation}
These are also the same expressions as in Ref.~\cite{bib: KYY}.
Thus the transformation parameters (\ref{eq: xis})-(\ref{eq:PQK}) are
found to agree with those in the superconformal tensor
calculus (Eqs.~(40), (42a), (42b) in \cite{bib: KUigc}).

At this moment, one question arises. 
When we go down from the conformal superspace to 
the component superconformal tensor calculus, the lowest components of
curved spinor gauge superfields are gauge-fixed to zero,
$h_{\ul\mu}{}^{\cal{A}'}|=0$, other than the vierbein.
This means that the lowest components of flat spinor gauge superfields
are also equal to zero, that is, $h_{\ul\mu}{}^{\cal{A}'}|
=\delta_{\ul\mu}{}^{\ul\alpha}h_{\ul\alpha}{}^{\cal{A}'}|=0$. 
However when we use the superspace version of the KU gauge, 
non-vanishing flat spinor gauge superfields $A_{\ul\alpha}$ and 
$f_{\ul{\alpha}}{}^A$ appear and 
correspond to the $A$ and $K_A$ compensations in the Poincar\'e
supersymmetry transformation in component approach as seen above.
Why does the component approach know non-vanishing spinor gauge
fields? What is the origin of them?

The answer is the resetting of gauge conditions. Let us consider
the $A$ gauge superfields as an example.
When going down to the component superconformal tensor calculus,
all the higher components of gauge parameter superfields 
other than the lowest are used to fix the superfield $A_{\ul\mu}$ and,
in particular, its lowest component 
$A_{\ul\mu}|=\delta_{\ul\mu}{}^{\ul\alpha}A_{\ul\alpha}|$ is 
set equal to zero by using the parameter $\der_\mu\xi(A)|$. 
On the other hand, when going down to the Poincar\'e superspace,
the whole gauge parameter superfields are used for $\xi(A)$, $\xi(D)$ 
and $\xi(K)$ to realize different gauge-fixing conditions. For example, 
$\xi(A)$ is chosen as $\xi(A)=\frac{3}{4i}\log(\b\Phi^0/\Phi^0)$ 
to have the KU gauge counterpart in the superspace~\cite{bib: KYY}. 
This form of gauge parameter superfield resets 
its component $\der_\mu\xi(A)|$ to a different value from the above,
and therefore $A_\mu|$ is generally reset to a 
nonzero quantity. In particular, the form of spinor $A$-gauge field should
coincide with the one appearing in the $A$-gauge compensation term in
the Poincar\'e supersymmetry transformation, 
since the supersymmetry transformation after gauge fixing is unique.

We show this resetting explicitly. When we perform 
the above $A$-gauge transformation with the parameter 
$\xi(A)=\frac{3}{4i}\log (\b\Phi^0/\Phi^0)$,
the spinor gauge field $A_\mu|$ is reset
from the initial value $A_\mu|=0$ as 
\begin{equation}
A_\mu| \,=\, 0 + \frac{3}{4i} \der_\mu\log (\b\Phi^0/\Phi^0)|
\,=\, \df{3i}{4}\frac{1}{\Phi^0}\delta_\mu{}^\alpha\na_\alpha\Phi^0\Big|.
\end{equation}
Plugging the expressions $\Phi^0=e^{G/6}$ and 
$\na_\alpha\Phi^0|=\frac{1}{3}e^{G/6}G_i\cal{D}^\Po_\alpha\Phi^i|$ 
in the KU gauge, we find 
\begin{equation}
A_\alpha|
=\df{i}{4}G_i\cal{D}^\Po_\alpha\Phi^i|, 
\end{equation}
which exactly reproduces $A_\alpha|$ used for the gauge parameter 
$\xi(A)'(\eta)$ in (\ref{eq: xis}). 
A similar discussion is possible for the $K_A$ gauge fields, 
but more complicated. What KU found is that 
one can avoid explicit and complicated computations 
of gauge transformations. The final expression of the spinor gauge
superfields can be found by identifying the Poincar\'e supersymmetry
transformation with the one which retains the $D$, $A$, $K_A$
gauge-fixing conditions.

\subsection{Vanishing superpotential case}

The gauge-fixing condition \eqref{eq:GF-K} for the case of 
vanishing superpotential is related to the formulation of
Ref.~\cite{bib: KKLVP} in component approach and also to
the isometric K\"ahler superspace of Ref.~\cite{bib: BG2}.
This can be seen in the same way as discussed before by using 
Eqs.~\eqref{eq:GF-K} and \eqref{eq:GF-K2}.
In particular, the transformation parameter of the resultant
Poincar\'e supersymmetry in Eq.~\eqref{eq:omQ} becomes 
\begin{equation}
\xi(A)'(\eta)
= \eta^{\ul\alpha}
\Big(A_{\ul\alpha}-\frac{i}{4}(F_{(a)}-\b{F}_{(a)})\cal{A}_{\ul\alpha}\Big) 
=\frac{i}4K_j \eta^\alpha\tilde{\cal{D}}_\alpha^{\rm P}\Phi^j 
-\frac{i}4K_{j*}\b{\eta}_{\d{\alpha}} 
\b{\tilde{\cal{D}}}^{{\rm P}\,\d{\alpha}}
\b{\Phi}^{j*}, 
\label{eq:omAGF-K}
\end{equation}
which is the same form as (\ref{eq: xis}) with a replacement $G\to K$. 
The other parameters $\xi(K)'(\eta)^A$ are 
also given by the same expressions as (\ref{eq:PQS}) and (\ref{eq:PQK}), 
with the same replacement $G\to K$ being performed. 

\bigskip

\section{Summary}
\label{sec: sum}

In this paper, we have discussed the YM matter coupled conformal SUGRA
in superspace and compare it with the component approach.
We have introduced the YM gauge superfield of internal symmetry in
conformal superspace by gauging the isometry of the K\"ahler manifold
spanned by chiral matter superfields.
The superconformal property of the gaugino superfield is derived by 
the Bianchi/Jacobi identities in Eq.~\eqref{eq: Wa}.
The YM gauge transformation laws of K\"ahler potential, superpotential
and compensator are studied from the superspace viewpoint in
Eqs.~\eqref{eq: FJ} and \eqref{eq: XaPhic}.

In section~\ref{sec:GF-G}, we have presented the superspace
gauge-fixing conditions leading to the canonically
normalized EH and RS terms, which conditions give the KU-gauge
counterpart in superspace. 
The relation between the Poincar\'e and conformal supersymmetry
transformations (the covariant derivatives) is discussed at superfield
level. In section~4 we have also shown the gauge-fixing procedure in
detail how the conformal superspace formulation is reduced to the
component approach (superconformal tensor calculus). 

In section~\ref{sec: corr}, 
the KU gauge in component approach is shown to be equivalent to 
the superspace gauge \eqref{eq:GF-G} written in terms of superfields.
Then the relations between the Poincar\'e and conformal supersymmetry
transformations are found to exactly correspond to each other
in superspace and component approaches.
Finally, we discuss several approaches with the canonically normalized
EH and RS terms in the conformal superspace viewpoint.

\subsection*{Acknowledgments}
\noindent
The authors thank Shuntaro Aoki, Tetsutaro Higaki, Tetsuji Kimura,
Michinobu Nishida, Yusuke Yamada and Naoki Yamamoto for useful 
discussions and comments.
The authors thank the Yukawa Institute for Theoretical Physics at Kyoto
University. The discussions during the YITP workshop YITP-W-15-20 on
``Microstructures of black holes'' were useful to complete this work. 
R.Y.\ is supported by Research Fellowships of Japan Society for
the Promotion of Science for Young Scientists Grant Number 16J03226.


\newpage

\begin{thebibliography}{99}
\bibitem{bib: KYY}
 T.~Kugo, R.~Yokokura and K.~Yoshioka,
  arXiv:1602.04441 [hep-th].
\bibitem{bib:KTVN}
  M.~Kaku, P.~K.~Townsend and P.~van Nieuwenhuizen,
  Phys.\ Rev.\ D {\bf 17} (1978) 3179.
\bibitem{bib:KT}
  M.~Kaku and P.~K.~Townsend,
  Phys.\ Lett.\ B {\bf 76} (1978) 54.
\bibitem{bib:TVN}
  P.~K.~Townsend and P.~van Nieuwenhuizen,
  Phys.\ Rev.\ D {\bf 19} (1979) 3166.
\bibitem{bib: CFGVP}
  E.~Cremmer, S.~Ferrara, L.~Girardello and A.~Van Proeyen,
  Nucl.\ Phys.\ B {\bf 212} (1983) 413.
\bibitem{Kugo:1982cu}
  T.~Kugo and S.~Uehara,
  Nucl.\ Phys.\ B {\bf 226} (1983) 49.
\bibitem{bib: KUigc} 
  T.~Kugo and S.~Uehara,
  Nucl.\ Phys.\ B {\bf 222} (1983) 125.
\bibitem{bib: KU} 
  T.~Kugo and S.~Uehara,
  Prog.\ Theor.\ Phys.\  {\bf 73} (1985) 235.
\bibitem{bib: KKLVP}
  R.~Kallosh, L.~Kofman, A.~D.~Linde and A.~Van Proeyen,
  Class.\ Quant.\ Grav.\  {\bf 17} (2000) 4269
   [Class.\ Quant.\ Grav.\  {\bf 21} (2004) 5017]
  [hep-th/0006179].
\bibitem{bib: B1} 
  D.~Butter,
  Annals Phys.\  {\bf 325} (2010) 1026
  [arXiv:0906.4399 [hep-th]].
\bibitem{bib: B2}
  D.~Butter,
  Nucl.\ Phys.\ B {\bf 828} (2010) 233
  [arXiv:0909.4901 [hep-th]].
\bibitem{Bagger:1982ab}
  J.~A.~Bagger,
  Nucl.\ Phys.\ B {\bf 211} (1983) 302.
\bibitem{bib: WB}
  J.~Wess and J.~Bagger,
  ``Supersymmetry and supergravity,''
  Princeton, USA: Univ. Pr. (1992) 259 p
\bibitem{bib: BG2}
  P.~Bin\'etruy, G.~Girardi and R.~Grimm,
  Phys.\ Rept.\  {\bf 343} (2001) 255
  [hep-th/0005225].
\bibitem{bib: BFI}
  D.~Butter,
  arXiv:1003.0249 [hep-th].
\bibitem{bib: FGKVP}
  S.~Ferrara, L.~Girardello, T.~Kugo and A.~Van Proeyen,
  Nucl.\ Phys.\ B {\bf 223} (1983) 191.
\bibitem{bib: SS}
  S.~J.~Gates, M.~T.~Grisaru, M.~Ro\v{c}ek and W.~Siegel,
  Front.\ Phys.\  {\bf 58} (1983) 1
  [hep-th/0108200].

\end{thebibliography}
\end{document}